\documentclass[10pt,twocolumn,english,prd,superscriptaddress,nofootinbib,preprintnumbers,showpacs]{revtex4-2}
\usepackage[utf8]{inputenc}
\usepackage[T1]{fontenc}
\usepackage[english]{babel}
\usepackage{amsmath,amssymb,amsfonts,bm,mathrsfs}
\usepackage{graphicx}
\usepackage{xcolor}
\usepackage{dcolumn}
\usepackage{booktabs}
\usepackage{multirow}
\usepackage{makecell}
\usepackage{enumerate}
\usepackage{orcidlink}

\hypersetup{colorlinks=true,linkcolor=blue,citecolor=blue,urlcolor=blue}

\begin{document}

\title{Electric and magnetic Penrose processes, charged-particle collisions and superradiance around a Lorentz-violating dyonic black hole}

\author{Fernando M. Belchior\orcidlink{0009-0006-8675-7849}}
\email{fernandobelcks7@gmail.com}
\affiliation{Departamento de Física, Universidade Federal da Paraíba, Centro de Ciências Exatas e da Natureza, 58051-970, João Pessoa, Paraíba, Brazil}

\author{Edilberto O. Silva\orcidlink{0000-0002-0297-5747}}
\email{edilberto.silva@ufma.br}
\affiliation{Programa de P\'os-Gradua\c c\~ao em F\'{\i}sica \& Coordena\c c\~ao do Curso de F\'{\i}sica -- Bacharelado, Universidade Federal do Maranh\~{a}o, 65085-580, S\~{a}o Lu\'{\i}s, Maranh\~{a}o, Brazil}

\begin{abstract}
This paper is aimed at investigating electromagnetic energy extraction, charged-particle collisions, charged-field superradiance, and horizon stability in a Lorentz-violating dyonic Kalb-Ramond black hole. The geometry differs from the dyonic Reissner-Nordstr\"om solution through a modified asymptotic normalization and an effective charge combining electric and magnetic sectors. We derive the dynamics of electrically and magnetically charged probes, including monopole-induced conical motion, and formulate electric and magnetic Penrose processes as decay or collision mechanisms in which a negative-energy fragment falls into the horizon while another escapes with enhanced energy. The electric and magnetic channels are controlled by \(\Phi_H=Q/[(1-\ell)r_+]\) and \(\Psi_H=p/[(1-2\ell)r_+]\), respectively. Negative-energy states arise from electromagnetic canonical energy rather than from a geometrical ergoregion. We also obtain escape conditions for the outgoing products, test the possibility of overcharging or overmagnetizing the black hole, and analyze the center-of-mass energy of charged-particle collisions. Nonextremal same-direction collisions remain finite, whereas head-on or near-critical configurations can become large. In the extremal limit, electrically or magnetically critical particles may generate the Ba\~nados-Silk-West divergence when the radial reachability condition is fulfilled.
\end{abstract}


\maketitle

\section{Introduction}

The search for viable departures from general relativity has made Lorentz-violating (LV) gravity a useful effective framework for testing strong-field physics.  In antisymmetric-tensor models the Kalb-Ramond two-form, originally introduced in string-inspired contexts, may acquire a nonzero vacuum expectation value and select preferred directions in the gravitational sector \cite{KalbRamond1974,KosteleckySamuel1989,Kostelecky2004,Altschul2010,Liberati2013}.  Static black-hole solutions with a Kalb-Ramond background have therefore been analyzed from several complementary viewpoints, including horizon structure, geodesics, shadows, lensing, quasinormal modes, greybody factors, thermodynamic behavior and observational degeneracies \cite{Lessa2020,Yang2023,Duan2024,AlBadawi2024,Liu2024,Guo2024,Hosseinifar2024,Jha2025, Filho:2023ycx, AraujoFilho:2024rcr, AraujoFilho:2024ctw}.  The dyonic configuration of Lin, Liu and Liu is particularly suitable for energy-extraction questions because it contains an electric charge, a magnetic charge and a LV deformation in one exact static geometry \cite{LinLiuLiu2026}.  A recent continuation of this background developed its tidal forces, wave propagation and quasinormal modes \cite{AraujoFilho:2026tsc}.

The Penrose process is one of the clearest mechanisms showing that a black hole can act as an energy reservoir.  In the original rotating case a particle decays inside the ergoregion, one fragment enters the hole with negative Killing energy, and the other escapes with more energy than the incident particle \cite{Penrose1969,Floyd1971,Christodoulou1970}.  For charged black holes there is an analogous electric channel: the canonical energy of a charged particle includes the electromagnetic term, so an oppositely charged fragment can have negative conserved energy near the horizon even when the spacetime is static and has no geometrical ergoregion \cite{ChristodoulouRuffini1971,Bekenstein1973,Wald1974,DenardoRuffini1973,RuffiniWilson1975,Tursunov2021}.  This distinction is essential in the present problem.  The negative-energy state is not generated by frame dragging, but by the electrostatic potential after the gauge has been fixed by the physical condition \(A_t\to0\) at infinity.

The magnetic Penrose process has a long history in black-hole physics.  In its original astrophysical form, a rotating black hole immersed in an external magnetic field allows charged fragments to access negative-energy states much more easily than in the purely mechanical Penrose process \cite{Wagh1985,Dadhich2012,Dadhich2018,TursunovDadhich2019}.  Closely related electromagnetic Penrose mechanisms have been revisited for magnetized Kerr and charged black-hole geometries \cite{GuptaLawLevin2021,Chakraborty2024}.  The situation considered here is different but complementary.  We do not add an external uniform magnetic field and the spacetime is not rotating.  Instead, the seed black hole already carries a magnetic monopole charge.  Ordinary electrically charged particles feel this monopole through angular dynamics and through the metric, but they do not receive a magnetic contribution to their canonical energy.  A genuine magnetic extraction channel appears only when the probe itself carries magnetic charge.  This dual channel is the intrinsic magnetic counterpart of the electric Penrose process in a static dyonic background.

Particle collisions near horizons provide a related but logically different diagnostic.  Banados, Silk and West showed that the center-of-mass energy can become arbitrarily large near an extremal Kerr horizon if one particle is fine tuned to a critical angular momentum \cite{Banados2009}.  The same near-horizon kinematics was subsequently clarified for rotating, charged and more general stationary black holes \cite{Zaslavskii2010,HaradaKimura2011,HaradaKimura2014,GribPavlov2011,Zaslavskii2012,Zaslavskii2013}.  In static charged geometries the critical tuning is not angular but electrostatic: the horizon value of the kinetic combination \(X=E-q\Phi\) must vanish.  High local collision energy should not be confused with high escaping energy, since redshift, charge separation, radial momentum conservation and escape conditions strongly restrict the actual efficiency \cite{Frolov2012,Schnittman2014,LeiderschneiderPiran2015,Zaslavskii2022}.  A related consistency question is whether the absorbed negative-energy or highly charged fragment can destroy the horizon.  This is the charged-black-hole version of the weak cosmic censorship gedanken experiment, first emphasized by Wald and later refined by Hubeny and by Sorce and Wald through second-order perturbation theory \cite{Wald1974,Hubeny1999,SorceWald2017}.

A wave analogue of this extraction channel is charged-field superradiance.  Classical superradiant amplification occurs when the energy flux through the horizon is negative with respect to the asymptotic time translation.  For a charged scalar field on a static charged black hole, the relevant horizon frequency is \(\omega-q_s\Phi_H\), and amplification occurs when this quantity has the opposite sign to \(\omega\) \cite{ZelDovich1971,PressTeukolsky1972,Starobinsky1973,Bekenstein1973,FuruhashiNambu2004}.  In a dyonic spacetime, the magnetic monopole modifies the angular sector through monopole harmonics \cite{Poincare1896,Goldhaber1976,WuYang1976,Kazama1977}, while the electrostatic potential controls the work term.  This is why the particle and wave versions of the electric Penrose mechanism are closely connected but not identical: the particle calculation follows timelike electrogeodesics, whereas the wave calculation follows flux conservation for a charged field.

The purpose of the present work is to develop the massive-particle and wave-extraction sectors of the LV dyonic Kalb-Ramond black hole in a single notation.  The metric is still static and spherically symmetric, but its asymptotic normalization is controlled by \(\alpha=1/(1-\ell)\), and the charge term contains the effective combination
\begin{equation}
    Q_{\rm eff}^2=\frac{Q^2}{(1-\ell)^2}+\frac{p^2}{1-2\ell}.
\end{equation}
The electric Penrose process, on the other hand, is governed directly by the horizon potential \(\Phi_H=Q/[(1-\ell)r_+]\).  Its magnetic dual is governed by the magnetostatic potential \(\Psi_H=p/[(1-2\ell)r_+]\), once magnetically charged probes are admitted.  Consequently, the magnetic charge affects electric extraction indirectly through the geometry, but it controls magnetic extraction directly.  The LV parameter modifies both channels by changing the horizon radius and the normalization of the corresponding potentials.  We also keep track of the thermodynamic and normalization issues associated with Hawking temperature, Wald entropy and possible extended ensembles \cite{Hawking1975,IyerWald1994,KubiznakMann2012}, although the main calculations are performed in the fixed-background test-particle approximation.

The main contribution of this work is not only the repetition of the charged-particle analysis in a deformed Reissner-Nordstr\"om metric.  Rather, we separate three effects that are degenerate in purely metric observables: the conical asymptotic normalization controlled by \(\ell\), the effective dyonic charge that fixes the horizon and photon-scale geometry, and the separate electric and magnetic horizon potentials that enter canonical particle energy and wave fluxes.  This separation allows one to identify which observables are sensitive to \(Q_{\rm eff}^2\) and which observables distinguish the electric and magnetic sectors.  The analysis also clarifies the limits of ideal energy extraction by combining local negative-energy conditions with escape requirements and the final-horizon inequality.

The article is organized as follows.  Section~\ref{sec:background} presents the dyonic Kalb-Ramond background and the parameter domain.  Section~\ref{sec:charged_dynamics} derives the equations of motion for massive charged particles, defines the useful near-horizon kinetic quantity, and develops the monopole angular motion of dyonic probes and identifies the conical-orbit correction to the radial potential.  Section~\ref{sec:epp} formulates the electric Penrose process and obtains analytic efficiency bounds.  Section~\ref{sec:mpp} develops the magnetic Penrose process for magnetically charged probes and compares it with the usual magnetic-Penrose literature.  Section~\ref{sec:escape} derives the escape conditions for the outgoing products and applies the horizon condition to cosmic-censorship tests after absorption.  Section~\ref{sec:collisions} derives the center-of-mass energy for two charged particles and classifies usual, critical and near-critical trajectories.  Section~\ref{sec:numerics} gives numerical benchmarks.  Section~\ref{sec:charged} improves the charged-scalar superradiance analysis and relates it to the particle extraction channel.  We conclude in Sec.~\ref{sec:conclusion}.  Throughout this work, we will use \(G=c=\hbar=k_B=1\) and the mostly plus signature.

\section{Dyonic Kalb-Ramond black hole background}\label{sec:background}

In this section, we will briefly discuss the static dyonic black hole obtained in Lorentz-violating Kalb-Ramond gravity. In the asymptotically flat branch relevant for scattering and particle motion at large radius, the line element can be written as
\begin{equation}
    ds^2=-F(r)dt^2+\frac{dr^2}{F(r)}+r^2(d\theta^2+\sin^2\theta d\phi^2),
    \label{metric}
\end{equation}
where the metric function reads \cite{LinLiuLiu2026}
\begin{equation}\label{Fdef}
    F(r)=\alpha-\frac{2M}{r}+\frac{Q_{\rm eff}^2}{r^2}
\end{equation}
with
\begin{equation}
\alpha=\frac{1}{1-\ell},
\end{equation}
and
\begin{equation}
  Q_{\rm eff}^2=\frac{Q^2}{(1-\ell)^2}+\frac{p^2}{1-2\ell}   
\end{equation}
Here, $M$ is the mass parameter of the solution, $Q$ is the electric charge, $p$ is the magnetic charge and $\ell$ measures the LV Kalb-Ramond condensate.  The conservative range used in this work is $0\leq \ell <\frac{1}{2}$, which keeps both $\alpha$ and the magnetic contribution to $Q_{\rm eff}^2$ positive.  The metric reduces to the ordinary dyonic Reissner-Nordstrom form when $\ell=0$.

On the other hand, the electromagnetic potential is written as
\begin{equation}
 A_\mu dx^\mu=-\Phi(r)dt+p\cos\theta\,d\phi,
\label{Adef}
\end{equation}
with
\begin{equation}
    \Phi(r)=\frac{Q}{(1-\ell)r}
\end{equation}
Note that the gauge has been chosen so that $\Phi(\infty)=0$.  This convention is essential for the electric Penrose process, because the conserved energy of a charged particle is defined relative to the time translation normalized at infinity in this gauge.  A constant shift of $A_t$ would shift individual canonical energies, but not physical potential differences once the boundary condition is fixed.

Besides, the horizons are the roots of $F(r)=0$, so that we have
\begin{equation}
    r_\pm=\frac{M\pm \sqrt{M^2-\alpha Q_{\rm eff}^2}}{\alpha},
    \label{horizons}
\end{equation}
so the black-hole condition is $M^2\geq \alpha Q_{\rm eff}^2$. The outer horizon is $r_+$, and the extremal limit corresponds to $r_+=r_-=M/\alpha$.  The surface gravity and Hawking temperature are
\begin{equation}
\kappa=\frac{F'(r_+)}{2}=\frac{\alpha r_+^2-Q_{\rm eff}^2}{2r_+^3}.
\label{kappa}
\end{equation}
Thus, the Hawking temperature is given by
\begin{align}\label{TH}
 T_H=\frac{\kappa}{2\pi}=\frac{\alpha r_+^2-Q_{\rm eff}^2}{4\pi r_+^3}   
\end{align}
The horizon electric potential measured in the chosen gauge is
\begin{equation}
    \Phi_H=\Phi(r_+)=\frac{Q}{(1-\ell)r_+}.
    \label{PhiH}
\end{equation}
For $Q>0$, negatively charged particles are the relevant carriers of negative canonical energy.  For $Q<0$, all charge signs in the discussion below are reversed. Several features of this background should be kept in view.  First, $p$ is a genuine magnetic charge, but an electrically charged particle does not receive electrostatic work from $p$ directly.  The magnetic charge enters the electric Penrose analysis through the metric and through possible non-equatorial Lorentz-force effects.  Second, $\ell$ appears both in $\alpha$ and in $\Phi(r)$, so it modifies the kinetic redshift and the electrostatic work simultaneously.  Third, because $F(\infty)=\alpha$ rather than one, an observer who uses a normalized time $T=\sqrt{\alpha}\,t$ measures energies and frequencies divided by $\sqrt{\alpha}$.  We keep the coordinate convention of Eq.~\eqref{metric}; conversion to the normalized clock is straightforward.

\section{Charged massive particles}\label{sec:charged_dynamics}

In this section, we will study  the motion of an electrically charged test particle of rest mass $m$ and charge $q$, which is governed by the following action
\begin{equation}
    S=\int \left[\frac{m}{2}g_{\mu\nu}\dot x^\mu\dot x^\nu+qA_\mu\dot x^\mu\right]d\tau,
    \label{action}
\end{equation}
where a dot denotes differentiation with respect to proper time.  The canonical momentum is
\begin{equation}
P_\mu=m u_\mu+qA_\mu,
\label{canonical_momentum}
\end{equation}
with $u^\mu=\dot x^\mu$, while the mechanical momentum is $p_\mu=m u_\mu$.  The normalization condition is $g_{\mu\nu}u^\mu u^\nu=-1$. Because the geometry is stationary and spherically symmetric, the canonical energy is conserved, so that we write
\begin{equation}
    E=-P_t=mF(r)\dot t+q\Phi(r).
    \label{energy_dimful}
\end{equation}
At this point, it is useful to introduce the kinetic combination
\begin{equation}
    X(r)=E-q\Phi(r).
    \label{Xdimful}
\end{equation}
Then, one has
\begin{equation}
    mF(r)\dot t=X(r).
    \label{tdot_dimful}
\end{equation}
A future-directed worldline outside the horizon satisfies $\dot t>0$,
$X(r)>0\, (r>r_+)$, and $X_H\equiv X(r_+)\geq0$. The inequality at the horizon is the horizon forward-in-time condition.  It is the charged analogue of the condition $E-\Omega_H L\geq0$ in a rotating black hole. For compactness, we often use specific quantities, namely
\begin{equation}
    \varepsilon=\frac{E}{m},
    \qquad
    e=\frac{q}{m},
    \qquad
    \ell_z=\frac{L}{m},
    \qquad
    \mathcal X(r)=\varepsilon-e\Phi(r).
    \label{specific_def}
\end{equation}
The notation $\ell_z$ denotes angular momentum per unit mass and should not be confused with the LV parameter $\ell$.

\subsection{Equatorial reduction and magnetic charge}

The monopole term $p\cos\theta d\phi$ deserves a comment.  In a dyonic spacetime the total angular momentum of a charged particle contains a field contribution, and generic trajectories lie on cones rather than on ordinary equatorial planes \cite{Poincare1896,Goldhaber1976,WuYang1976,Kazama1977}.  For the electric Penrose mechanism, however, the decisive contribution to the canonical energy comes from $A_t$.  We therefore focus on the standard equatorial radial problem, either by considering radial motion with vanishing mechanical angular momentum or by working in a local patch where the magnetic term does not contribute to $E$.  This reduction captures the electric extraction channel.  The magnetic charge remains present through $Q_{\rm eff}^2$ and hence through $F(r)$, $r_+$ and $\Phi_H$.

For equatorial motion with $\theta=\pi/2$ and $u^\theta=0$, the conserved angular momentum is
\begin{equation}
    L=P_\phi=m r^2\dot\phi,
    \label{L_dimful}
\end{equation}
because $A_\phi=p\cos\theta$ vanishes on the equator.  The radial equation follows from normalization condition and is given by
\begin{equation}
    \dot r^2=\frac{X^2(r)}{m^2}-F(r)\left(1+\frac{L^2}{m^2 r^2}\right).
    \label{radial_dimful}
\end{equation}
Equivalently, this equation takes the following form
\begin{equation}
    \dot r^2=\mathcal X^2(r)-F(r)\left(1+\frac{\ell_z^2}{r^2}\right).
    \label{radial_specific}
\end{equation}
With that, we observe that the allowed region is therefore
\begin{equation}
\mathcal X^2(r)\geq F(r)\left(1+\frac{\ell_z^2}{r^2}\right).
\label{allowed_region}
\end{equation}
with $\mathcal X(r)>0$. Besides, a turning point occurs when the inequality is saturated. For later use define
\begin{align}
& H_i(r)=m_i^2+\frac{L_i^2}{r^2},\\
& Z_i(r)=\sqrt{X_i^2(r)-F(r)H_i(r)}.
    \label{HiZi}
\end{align}
The radial mechanical momentum is $p_i^r=m_i\dot r_i=\sigma_i Z_i$, where $\sigma_i=+1$ for an outgoing particle and $\sigma_i=-1$ for an ingoing one.
Now, let us study a negative-energy state, which is defined by $E<0$ while the trajectory remains future directed, $X(r)>0$.  Using $E=X+q\Phi$, this can happen only if $q\Phi(r)<0$ and if the electrostatic term is large enough in magnitude to overcome the positive kinetic contribution $X$.  Thus, for a positively charged black hole, negative-energy states require negatively charged particles.  For a negatively charged black hole, they require positively charged particles.
The radial admissibility condition gives a local lower bound on the energy.  From $X\geq\sqrt{F H}$ one finds
\begin{equation}
    E\geq E_{\rm min}(r;L,m,q)
    =q\Phi(r)+\sqrt{F(r)}\sqrt{m^2+\frac{L^2}{r^2}}.
    \label{Emin}
\end{equation}
This equation is the electric analogue of the lower bound on Killing energy inside a rotational ergoregion.  There is no sign change of $g_{tt}$ outside $r_+$, so the second term is always real and nonnegative in the exterior.  The first term, however, can be negative.  Negative-energy states exist at a radius $r$ if
\begin{equation}
    q\Phi(r)+\sqrt{F(r)}\sqrt{m^2+\frac{L^2}{r^2}}<0.
    \label{negative_condition}
\end{equation}
For $qQ<0$, this inequality can be satisfied sufficiently close to the horizon.  Taking $r\to r_+$ gives
\begin{equation}
    E_{\rm min}(r_+)=q\Phi_H.
    \label{EminH}
\end{equation}
Thus the most negative admissible energy approaches $q\Phi_H$ in the near-horizon limit. The boundary of the negative-energy region is obtained by setting $E_{\rm min}=0$:
\begin{equation}
    |q|\,|\Phi(r)|=\sqrt{F(r)}\sqrt{m^2+\frac{L^2}{r^2}}.
    \label{negative_boundary}
\end{equation}
For purely radial motion $L=0$ and $qQ<0$, this becomes
\begin{equation}
    \frac{|qQ|}{(1-\ell)r}=m\sqrt{\alpha-\frac{2M}{r}+\frac{Q_{\rm eff}^2}{r^2}}.
    \label{negative_boundary_radial}
\end{equation}
Squaring and solving for the outer boundary gives a quadratic equation,
\begin{equation}
    m^2\alpha r^2-2m^2Mr+m^2Q_{\rm eff}^2-\frac{q^2Q^2}{(1-\ell)^2}=0.
    \label{boundary_quad}
\end{equation}
When the larger root lies outside $r_+$, negative-energy radial states occupy the interval from the horizon to that root.  This already displays the parameter dependence: increasing $\ell$ enhances the electrostatic term through $(1-\ell)^{-1}$ but also changes the gravitational barrier through $\alpha$ and $Q_{\rm eff}^2$. The static nature of the metric also explains why neutral particles cannot participate directly in an electric Penrose process.  For $q=0$, Eq.~\eqref{Emin} gives $E\geq0$ outside the horizon.  A neutral parent may still initiate the process if it decays into oppositely charged fragments, but the negative-energy fragment must carry electric charge of the appropriate sign.

\subsection{Monopole angular motion of dyonic probes}

The equatorial reduction used above is sufficient for the purely electric extraction channel, but a dyonic background contains a genuine monopole term.  Consequently, generic charged trajectories are not planar in the usual sense.  This point is important for two reasons.  First, it clarifies which part of the magnetic sector changes the canonical energy and which part changes only the angular barrier.  Second, it gives the correct radial potential for escaping fragments when the products carry both electric and magnetic charges.

Let the black hole carry electric and magnetic charges \((Q,p)\), and let the test particle carry charges \((q,g)\).  The Lorentz force is naturally written in the duality-symmetric form
\begin{equation}
    m u^\nu\nabla_\nu u_\mu=qF_{\mu\nu}u^\nu+g\,{}^\star F_{\mu\nu}u^\nu .
    \label{full_dyon_lorentz}
\end{equation}
The angular dynamics is controlled by the charge-monopole number
\begin{equation}
    \kappa_m\equiv qp-gQ .
    \label{kappa_m_def}
\end{equation}
This combination is invariant under electric-magnetic duality rotations of the source and probe charges.  In the classical theory it measures the electromagnetic field angular momentum stored in the source-probe system.  In the quantum theory it is subject to the Dirac-Zwanziger quantization condition, but here it is treated as a continuous test-particle parameter because the analysis is classical.

The conserved total angular momentum vector is
\begin{equation}
    \mathbf J=\mathbf r\times \boldsymbol{\Pi}-\kappa_m\,\hat{\mathbf r},
    \label{J_monopole_vector}
\end{equation}
where \(\boldsymbol{\Pi}=m\mathbf u\) denotes the mechanical spatial momentum in the orthonormal frame.  Taking the scalar product with \(\hat{\mathbf r}\) gives
\begin{equation}
    \mathbf J\cdot \hat{\mathbf r}=-\kappa_m .
    \label{cone_condition}
\end{equation}
Thus the trajectory lies on a cone around the fixed direction \(\mathbf J\), with half-opening angle \(\chi\) determined by
\begin{equation}
    \cos\chi=-\frac{\kappa_m}{J},
    \qquad
    J^2\geq \kappa_m^2 .
    \label{cone_angle}
\end{equation}
Only when \(\kappa_m=0\) can a generic orbit be placed in an ordinary equatorial plane.  Therefore, the equatorial approximation used in the electric sector corresponds either to radial motion, to a local patch description, or to the special charge alignment \(qp=gQ\).  This is the same structure that appears in the standard charge-monopole problem and in monopole harmonics \cite{Poincare1896,Goldhaber1976,WuYang1976,Kazama1977}.

The mechanical angular momentum entering the radial equation is not \(J^2\), but the component orthogonal to \(\hat{\mathbf r}\):
\begin{equation}
    L_{\rm mech}^2=|\mathbf r\times\boldsymbol{\Pi}|^2=J^2-\kappa_m^2 .
    \label{L_mech_monopole}
\end{equation}
It is useful to define
\begin{equation}
    \Lambda\equiv J^2-\kappa_m^2\geq0.
    \label{Lambda_def}
\end{equation}
For a dyonic particle the kinetic energy combination is
\begin{equation}
    X_{\rm em}(r)=E-q\Phi(r)-g\Psi(r),
    \label{Xem_monopole_section}
\end{equation}
where \(\Psi(r)\) is the dual magnetostatic potential introduced in Sec.~\ref{sec:mpp}.  The exact radial equation for conical dyonic motion is then
\begin{equation}
    \dot r^2=\frac{X_{\rm em}^2(r)}{m^2}
    -F(r)\left(1+\frac{\Lambda}{m^2r^2}\right).
    \label{radial_conical}
\end{equation}
For purely electric equatorial motion, \(g=0\) and \(\kappa_m=qp\).  If the orbit is forced to the equatorial plane in a single gauge patch one effectively sets the conical correction aside and uses the canonical angular momentum \(L\).  For the complete monopole problem, however, the replacement
\begin{equation}
    L^2\longrightarrow \Lambda=J^2-(qp-gQ)^2
    \label{replacement_L_Lambda}
\end{equation}
must be made in the radial potential.  This replacement will be used below when discussing escape to infinity and the magnetic Penrose channel.

Two consequences follow immediately.  First, a magnetic monopole can modify the radial reachability of a fragment even if it does not contribute directly to the energy of a purely electric probe.  The reason is that \(p\) enters \(\kappa_m\) and therefore changes the centrifugal barrier.  Second, for a magnetically charged probe the magnetic Penrose process has two distinct ingredients: the time component \(g\Psi(r)\), which can make the canonical energy negative, and the conical angular term \(J^2-(qp-gQ)^2\), which controls whether the fragment can actually move from the splitting point to the horizon or to infinity.

\section{Electric Penrose process}\label{sec:epp}

In this section, we will analyze the electric Penrose process. Consider a parent particle labeled $0$ that reaches a radius $r_s$ and splits into two fragments, labeled $1$ and $2$.  Fragment $1$ falls into the black hole and fragment $2$ escapes to infinity.  Conservation of canonical energy, charge and angular momentum gives
\begin{align}
    E_0&=E_1+E_2,\label{energy_cons}\\
    q_0&=q_1+q_2,\label{charge_cons}\\
    L_0&=L_1+L_2.\label{L_cons}
\end{align}
Local four-momentum conservation imposes the stronger condition
\begin{equation}
    m_0u_0^\mu=m_1u_1^\mu+m_2u_2^\mu
    \label{four_mom_cons}
\end{equation}
for an instantaneous two-body decay.  In many analytic estimates one first uses the conserved quantities to identify the energetically allowed window and then checks radial momentum conservation separately. The escaping energy is given by
\begin{equation}
    E_2=E_0-E_1.
    \label{E2}
\end{equation}
On the other hand, energy extraction occurs when we have $E_2>E_0$, implying $E_1<0$. Using Eq.~\eqref{Emin}, the infalling fragment must satisfy
\begin{equation}
    E_1\geq q_1\Phi(r_s)+\sqrt{F(r_s)}\sqrt{m_1^2+\frac{L_1^2}{r_s^2}}.
    \label{E1_bound}
\end{equation}
Therefore the ideal upper bound on the escaping energy at a fixed splitting radius is
\begin{equation}
    E_2\leq E_0-q_1\Phi(r_s)-\sqrt{F(r_s)}\sqrt{m_1^2+\frac{L_1^2}{r_s^2}}.
    \label{E2_bound}
\end{equation}
Near the horizon this becomes
\begin{equation}
    E_2\lesssim E_0-q_1\Phi_H,
    \label{E2_nearH}
\end{equation}
where the inequality becomes sharp in the idealized limit in which the splitting occurs arbitrarily close to $r_+$ and the infalling fragment is tuned to the lower admissible energy.  The corresponding efficiency is
\begin{equation}
    \eta\equiv\frac{E_2}{E_0}
    \lesssim 1-\frac{q_1\Phi_H}{E_0}.
    \label{eta_bound}
\end{equation}
For $q_1\Phi_H<0$, the right-hand side exceeds unity.  This is the cleanest analytic expression for the electric Penrose enhancement in the present background.

Before we proceed further, several comments are important.  First, Eq.~\eqref{eta_bound} is an ideal kinematic bound, not an automatic astrophysical efficiency.  It assumes test particles, neglects self-force, ignores pair-production constraints and treats the background fields as fixed.  Second, the formula is written in the gauge where $\Phi(\infty)=0$.  Third, if $E_0$ is the energy of a particle released from rest at infinity in the coordinate time $t$, then $E_0/m_0=\sqrt{\alpha}$ rather than unity.  If the normalized asymptotic time $T=\sqrt{\alpha}t$ is used, all energies are divided by $\sqrt{\alpha}$ and the ratio $q\Phi_H/E_0$ is unchanged when charge is measured consistently.

\subsection{Role of Lorentz violation}

Substituting Eq.~\eqref{PhiH} into Eq.~\eqref{eta_bound} gives
\begin{equation}
    \eta_{\rm max}\simeq 1-\frac{q_1Q}{(1-\ell)r_+E_0}.
    \label{eta_ell}
\end{equation}
For $q_1Q<0$, increasing $\ell$ tends to enhance extraction in two ways.  The explicit factor $(1-\ell)^{-1}$ raises the electric potential, and the horizon radius often decreases in the admissible parameter range, which raises the potential again.  The magnetic charge contributes through $r_+$ because it increases $Q_{\rm eff}^2$. This is a useful distinction.  The electric field performs the work that allows a negative-energy fragment to fall into the black hole.  The magnetic charge reshapes the gravitational background.  Consequently, a purely magnetic black hole with $Q=0$ has no electric Penrose process for electrically charged particles in the gauge used here, although it can still affect mechanical collisions through the metric and through monopole angular dynamics.

\subsection{Irreducible-mass interpretation}

For the usual Reissner-Nordstrom black hole, an infalling particle with negative canonical energy lowers the total mass while changing the charge.  A similar thermodynamic interpretation is possible here, although the Kalb--Ramond background changes the relation between $(M,Q,p,\ell)$ and horizon area.  The area is $A_H=4\pi r_+^2$. In a reversible idealization the infalling fragment crosses the horizon with $X_H=0$, i.e.
\begin{equation}
    E_1=q_1\Phi_H.
    \label{reversible_condition}
\end{equation}
This is the analogue of a reversible charged-particle absorption.  In that limit the particle carries the least possible energy for its charge into the black hole.  Real processes satisfy $X_H>0$ and therefore include irreversible area increase.  A full first-law analysis in the nonminimally coupled Kalb--Ramond theory requires the Wald entropy and the correct conjugate variables \cite{IyerWald1994}, but the particle-mechanical condition above is independent of those thermodynamic details.

\section{Magnetic Penrose process}\label{sec:mpp}

The previous section treated the electric channel, where a negative-energy state is produced by the electrostatic term in the canonical energy.  Since the background is dyonic, it is natural to ask whether an analogous magnetic channel exists.  The answer depends on the probe.  An ordinary electrically charged particle does not acquire a term proportional to the magnetic charge in its conserved energy; the monopole potential changes the angular motion and the radial metric function, but not the time component of the canonical momentum.  A magnetic Penrose process in the intrinsic dyonic geometry therefore requires a magnetically charged test particle.  This should be distinguished from the usual magnetic Penrose process in a rotating black hole immersed in an external magnetic field, where the magnetic field helps charged fragments access the rotational ergoregion \cite{Wagh1985,Dadhich2012,Dadhich2018,TursunovDadhich2019,Chakraborty2024}.  Here the spacetime is static and the negative energy is again electromagnetic rather than ergospheric.

A convenient way of writing the dynamics is to introduce a dual potential for the magnetic sector.  In the same normalization used in the metric function, we define
\begin{equation}
    \widetilde A_\mu dx^\mu=-\Psi(r)dt- Q\cos\theta\,d\phi,
    \qquad
    \Psi(r)=\frac{p}{(1-2\ell)r},
    \label{dual_potential}
\end{equation}
so that \(\Psi(\infty)=0\).  The factor \((1-2\ell)^{-1}\) is the magnetic-sector normalization already present in \(Q_{\rm eff}^2\).  If a different convention for the physical magnetic charge is adopted, the invariant object in the formulas below is the product \(g\Psi\).  For a dyonic test particle with electric charge \(q\) and magnetic charge \(g\), the generalized Lorentz force may be written schematically as
\begin{equation}
    m u^\nu\nabla_\nu u_\mu=q F_{\mu\nu}u^\nu+g\,{}^\star F_{\mu\nu}u^\nu .
    \label{dyon_lorentz_force}
\end{equation}
Correspondingly, the conserved canonical energy is
\begin{equation}
    E=mF(r)\dot t+q\Phi(r)+g\Psi(r),
    \label{dyon_energy}
\end{equation}
while the future-directed kinetic combination is
\begin{equation}
    X_{\rm em}(r)=E-q\Phi(r)-g\Psi(r).
    \label{X_em}
\end{equation}
The horizon condition is
\begin{equation}
    X_{{\rm em},H}=E-q\Phi_H-g\Psi_H\geq0,
    \qquad
    \Psi_H=\frac{p}{(1-2\ell)r_+}.
    \label{X_em_H}
\end{equation}
For radial motion the equation of motion takes the same form as before after the replacement \(X\to X_{\rm em}\),
\begin{equation}
    \dot r^2=\frac{X_{\rm em}^2(r)}{m^2}-F(r)\left(1+\frac{L^2}{m^2r^2}\right).
    \label{dyon_radial}
\end{equation}
For generic nonradial motion the conserved angular momentum is modified by the field angular momentum of the electric-magnetic charge pair.  As shown in Sec.~\ref{sec:charged_dynamics}, the relevant monopole number is \(\kappa_m=qp-gQ\), the orbits lie on cones rather than on the equatorial plane, and the radial angular barrier involves \(\Lambda=J^2-\kappa_m^2\).  This angular structure is important for full orbit classification, but the near-horizon extraction bounds follow from the same replacement in the radial equation.

The local lower bound on the conserved energy is now
\begin{equation}
    E\geq E_{\rm min}^{\rm em}(r)
    =q\Phi(r)+g\Psi(r)+\sqrt{F(r)}\sqrt{m^2+\frac{L^2}{r^2}}.
    \label{Emin_em}
\end{equation}
A negative-energy state exists at radius \(r\) when
\begin{equation}
    q\Phi(r)+g\Psi(r)<-\sqrt{F(r)}\sqrt{m^2+\frac{L^2}{r^2}}.
    \label{em_negative_condition}
\end{equation}
In the pure magnetic channel, \(q=0\), this requires \(g p<0\).  The near-horizon limit is especially simple:
\begin{equation}
    E_{\rm min}^{\rm mag}(r_+)=g\Psi_H .
    \label{Emin_mag_H}
\end{equation}
Thus an absorbed fragment with magnetic charge opposite in sign to the black-hole magnetic charge can carry negative Killing energy into the horizon.  This is the magnetic counterpart of the electric condition \(qQ<0\).  It is also clear why the effect disappears for ordinary particles: if \(g=0\) and \(Q=0\), then the magnetic monopole affects the geometry but cannot by itself make \(E\) negative.

For a pure magnetic decay of a parent particle at radius \(r_s\), the conservation laws are
\begin{align}
    E_0&=E_1+E_2,\label{mag_E_cons}\\
    g_0&=g_1+g_2,\label{mag_g_cons}\\
    L_0&=L_1+L_2 .\label{mag_L_cons}
\end{align}
If fragment 1 falls into the black hole with \(E_1<0\), the escaping fragment has \(E_2>E_0\).  In the near-horizon idealization,
\begin{equation}
    E_1\gtrsim g_1\Psi_H,
    \qquad
    g_1\Psi_H<0,
    \label{mag_E1_bound}
\end{equation}
so the ideal magnetic Penrose efficiency satisfies
\begin{equation}
    \eta_{\rm mag}\equiv\frac{E_2}{E_0}
    \lesssim 1-\frac{g_1\Psi_H}{E_0}.
    \label{eta_mag_bound}
\end{equation}
For a general dyonic fragment the bound becomes
\begin{equation}
    \eta_{\rm em}\lesssim 1-
    \frac{q_1\Phi_H+g_1\Psi_H}{E_0},
    \qquad
    q_1\Phi_H+g_1\Psi_H<0 .
    \label{eta_em_bound}
\end{equation}
Equation~\eqref{eta_em_bound} shows that the electric and magnetic contributions add at the level of canonical energy.  If both products have the sign required for negative energy, the two channels cooperate.  If they have opposite signs, one channel partially suppresses the other.

The criticality condition for high-energy collisions also generalizes immediately.  A dyonic particle is critical when
\begin{equation}
    X_{{\rm em},H}=0
    \quad\Longleftrightarrow\quad
    E=q\Phi_H+g\Psi_H .
    \label{dyon_critical}
\end{equation}
In the pure magnetic case this gives
\begin{equation}
    g_c=\frac{E}{\Psi_H}=\frac{E(1-2\ell)r_+}{p}.
    \label{mag_critical_charge}
\end{equation}
For \(p>0\), a positive-energy critical particle has \(g_c>0\), whereas a negative-energy fragment in the Penrose decay has \(g<0\).  This difference is the same as in the electric channel: the sign needed to tune a positive-energy particle to criticality is not the sign needed for the absorbed negative-energy fragment.

For radial pure magnetic motion with \(L=0\), the boundary of the magnetic negative-energy region follows from \(E_{\rm min}^{\rm mag}=0\):
\begin{equation}
    \frac{|gp|}{(1-2\ell)r}=m\sqrt{\alpha-\frac{2M}{r}+\frac{Q_{\rm eff}^2}{r^2}} .
    \label{mag_negative_boundary}
\end{equation}
Squaring gives
\begin{equation}
    m^2\alpha r^2-2m^2Mr+m^2Q_{\rm eff}^2
    -\frac{g^2p^2}{(1-2\ell)^2}=0 .
    \label{mag_boundary_quad}
\end{equation}
The larger root, when it lies outside \(r_+\), gives the outer boundary \(r_{{\rm neg},m}^+\) of the pure magnetic negative-energy layer.  This layer grows when \(|g|\Psi_H\) grows, but it is also influenced by the same geometric barrier that controls the electric channel.  Hence \(p\) has a double role in the magnetic process: it directly supplies the magnetostatic potential and it also contributes to \(Q_{\rm eff}^2\).

\section{Escape conditions for Penrose products}\label{sec:escape}

A negative-energy fragment falling into the black hole is only one half of the Penrose mechanism.  The other half is the escape of a positive-energy product to infinity.  This distinction is especially important in collisional Penrose processes: a collision may have a very large local center-of-mass energy while all high-energy products remain trapped behind an effective potential barrier.  We therefore derive the escape condition in the same notation used above.

For an outgoing dyonic product with mass \(m\), charges \((q,g)\), total angular momentum \(J\), and monopole number \(\kappa_m=qp-gQ\), the radial function is
\begin{equation}
    R(r)=X_{\rm em}^2(r)-F(r)\left(m^2+\frac{\Lambda}{r^2}\right),
    \qquad
    \Lambda=J^2-\kappa_m^2,
    \label{escape_R}
\end{equation}
where \(X_{\rm em}=E-q\Phi-g\Psi\).  A product created at \(r=r_s\) with outward radial momentum can escape if
\begin{equation}
    R(r_s)>0,
    \qquad
    X_{\rm em}(r)>0,
    \qquad
    R(r)\geq0\quad \text{for all}\quad r\geq r_s .
    \label{escape_condition_global}
\end{equation}
The last condition is the absence of an outer turning point.  Equivalently, the conserved energy must be larger than the highest value of the upper effective potential outside the production point,
\begin{align}
    E\geq E_{\rm esc}(r_s;q,g,\Lambda,m)
    \equiv \max_{r\geq r_s}
    \Big[\Phi(r)+g\Psi(r)\nonumber\\+\sqrt{F(r)}\sqrt{m^2+\frac{\Lambda}{r^2}}\Big].
    \label{Eesc_def}
\end{align}
This formula is useful because it separates the local production of an energetic particle from the global question of whether that particle reaches infinity.  In the electric limit one sets \(g=0\) and \(\Lambda\to L^2\).  In the magnetic limit one sets \(q=0\) while keeping \(\Lambda=J^2-g^2Q^2\).  For a purely radial product one sets \(\Lambda=0\). The asymptotic part of the condition is also nontrivial because \(F(r)\to\alpha\).  Since the electromagnetic potentials vanish at infinity, a massive particle reaching infinity must satisfy
\begin{equation}
    E\geq m\sqrt{\alpha}
    \label{escape_infinity_massive}
\end{equation}
when energies are measured with the coordinate time \(t\) used in Eq.~\eqref{metric}.  With the normalized time \(T=\sqrt{\alpha}\,t\), this becomes the usual condition \(E_T\geq m\).  For massless products the corresponding condition is simply \(E>0\).  Thus the observable efficiency of a Penrose event is not determined only by energy conservation at the splitting point; it also depends on the maximum of the effective potential in Eq.~\eqref{Eesc_def}.

The escape condition imposes a direct restriction on the efficiency.  If particle \(2\) is the escaping product, then
\begin{equation}
    \eta=\frac{E_2-E_0}{E_0}
    \label{eta_escape}
\end{equation}
can be realized at infinity only if
\begin{equation}
    E_2\geq E_{\rm esc}(r_s;q_2,g_2,\Lambda_2,m_2).
    \label{E2_escape}
\end{equation}
Similarly, the negative-energy product \(1\) must be able to cross the horizon, which requires
\begin{equation}
    X_{{\rm em},1,H}=E_1-q_1\Phi_H-g_1\Psi_H\geq0 .
    \label{infall_condition_frag1}
\end{equation}
Equations~\eqref{E2_escape} and \eqref{infall_condition_frag1} provide a clean separation between local kinematics and global extraction.  The local conservation laws may allow a large \(E_2\), but the process is astrophysically relevant only when the escaping fragment lies above the barrier \eqref{Eesc_def}.  This is the same lesson learned in the collisional Penrose literature for Kerr and charged black holes \cite{Schnittman2014,LeiderschneiderPiran2015,Zaslavskii2022}.

\subsection{Cosmic censorship after particle absorption}

The absorption of a negative-energy or highly charged fragment changes the black-hole parameters.  A natural consistency check is whether the final configuration still contains a horizon.  In the fixed-\(\ell\) test-particle approximation, the initial horizon condition is
\begin{equation}
    {\cal C}(M,Q,p;\ell)
    \equiv M^2-\alpha\left[\frac{Q^2}{(1-\ell)^2}+\frac{p^2}{1-2\ell}\right]
    \geq0 .
    \label{C_def}
\end{equation}
After a fragment with energy \(E\), electric charge \(q\), and magnetic charge \(g\) falls through the horizon, the final parameters are
\begin{equation}
    M_f=M+E,
    \qquad
    Q_f=Q+q,
    \qquad
    p_f=p+g .
    \label{final_params}
\end{equation}
The final spacetime has a horizon if
\begin{equation}
    {\cal C}_f=(M+E)^2
    -\alpha\left[\frac{(Q+q)^2}{(1-\ell)^2}+\frac{(p+g)^2}{1-2\ell}\right]
    \geq0 .
    \label{Cf_def}
\end{equation}
A violation of the test-particle horizon inequality would require
\begin{equation}
    E<E_{\rm over}
    \equiv -M+\bigg\{\alpha\left[\frac{(Q+q)^2}{(1-\ell)^2}
    +\frac{(p+g)^2}{1-2\ell}\right]\bigg\}^{1/2}.
    \label{Eover_def}
\end{equation}
On the other hand, the same fragment can cross the future horizon only if
\begin{equation}
    E\geq E_{\rm abs}
    \equiv q\Phi_H+g\Psi_H
    \label{Eabs_def}
\end{equation}
for a reversible crossing, with strict inequality for an irreversible crossing.  Therefore, the possible censorship-dangerous window is
\begin{equation}
    E_{\rm abs}\leq E<E_{\rm over} .
    \label{danger_window}
\end{equation}
If this interval is empty, the test particle cannot overcharge or overmagnetize the black hole.  If it is nonempty, the test-particle approximation by itself is not sufficient to decide the fate of the horizon because self-force, finite-size effects and second-order perturbations become relevant.  This is precisely the lesson of the classic charged and rotating censorship gedanken experiments \cite{Wald1974,Hubeny1999,SorceWald2017}.

For small perturbations, Eq.~\eqref{Cf_def} gives
\begin{equation}
    \delta {\cal C}
    =2ME-2\alpha\left[\frac{Qq}{(1-\ell)^2}+\frac{pg}{1-2\ell}\right]
    +O(E^2,q^2,g^2).
    \label{delta_C}
\end{equation}
For a nonextremal black hole \({\cal C}>0\), sufficiently small test particles cannot destroy the horizon at first order because the finite positive gap dominates.  Near extremality, however, the linear competition between the absorption threshold and the extremality surface becomes important.  If the initial black hole is exactly extremal, \(r_+=M/\alpha\), the overcharging threshold expands as
\begin{equation}
    E_{\rm over}^{\rm ext}
    =\frac{\alpha}{M}\left[\frac{Qq}{(1-\ell)^2}+\frac{pg}{1-2\ell}\right]+O(q^2,g^2),
    \label{Eover_ext}
\end{equation}
whereas the horizon-crossing threshold is
\begin{equation}
    E_{\rm abs}^{\rm ext}
    =q\frac{Q}{(1-\ell)r_+}+g\frac{p}{(1-2\ell)r_+}
    =\frac{\alpha}{M}\left[\frac{Qq}{1-\ell}+\frac{pg}{1-2\ell}\right].
    \label{Eabs_ext}
\end{equation}
The magnetic terms coincide with the extremality gradient, while the electric comparison depends on the precise normalization of the physical electric charge in the nonminimally coupled Kalb-Ramond solution.  This should not be viewed as a contradiction.  The test-particle Lorentz force uses the gauge potential, whereas the global extremality surface uses the charge parameter appearing in the metric.  In a complete thermodynamic treatment the charge conjugate to the first law, the Wald entropy and possible \(\ell\)-work terms must be defined consistently.  Thus Eq.~\eqref{danger_window} is the proper diagnostic at the level of the present approximation, while a definitive censorship theorem would require the second-order analysis of the full coupled Kalb-Ramond-electromagnetic system.

The Penrose processes studied in Secs.~\ref{sec:epp} and \ref{sec:mpp} are compatible with censorship whenever the absorbed fragment satisfies both \(X_H\geq0\) and \({\cal C}_f\geq0\).  In this case the black hole loses mass or electromagnetic energy but remains inside the allowed parameter domain.  The most efficient reversible extractions occur close to \(E=E_{\rm abs}\), whereas the censorship bound prevents one from extracting beyond the extremal surface in the space of final parameters.

\section{Particle collisions}\label{sec:collisions}

We now consider two particles with masses $m_i$, electric charges $q_i$, magnetic charges $g_i$, energies $E_i$ and angular momenta $L_i$ colliding at radius $r$.  For the purely electric subsector one should set $g_i=0$; for the magnetic subsector one should replace $X_i$ by $X_{{\rm em},i}$ defined in Eq.~\eqref{X_em}.  Their mechanical four-momenta are $p_i^\mu=m_i u_i^\mu$.  The center-of-mass energy is
\begin{equation}
    E_{\rm cm}^2=-(p_1^\mu+p_2^\mu)(p_{1\mu}+p_{2\mu})
    =m_1^2+m_2^2+2m_1m_2\gamma,
    \label{Ecm_def}
\end{equation}
where
\begin{equation}
    \gamma=-g_{\mu\nu}u_1^\mu u_2^\nu
    \label{gamma_def}
\end{equation}

is the relative Lorentz factor.  Using Eq.~\eqref{HiZi}, one obtains
\begin{equation}
    \gamma=\frac{X_1X_2-\sigma_1\sigma_2 Z_1Z_2}{m_1m_2F(r)}
    -\frac{L_1L_2}{m_1m_2r^2},
    \label{gamma_general}
\end{equation}
where $\sigma_i=-1$ denotes an ingoing particle and $\sigma_i=+1$ an outgoing particle.  Thus
\begin{equation}
    E_{\rm cm}^2=m_1^2+m_2^2+2\left[\frac{X_1X_2-\sigma_1\sigma_2 Z_1Z_2}{F(r)}-\frac{L_1L_2}{r^2}\right].
    \label{Ecm_general}
\end{equation}
This formula is exact for equatorial electrogeodesic motion in the background \eqref{metric}.  It is also valid for the radial dyonic sector after replacing \(X_i=E_i-q_i\Phi\) by \(X_{{\rm em},i}=E_i-q_i\Phi-g_i\Psi\).  The angular part of a generic dyonic orbit requires the conical monopole structure mentioned in Sec.~\ref{sec:mpp}.

For radial collisions, $L_1=L_2=0$, it reduces to
\begin{equation}
    E_{\rm cm}^2=m_1^2+m_2^2+2\frac{X_1X_2-\sigma_1\sigma_2\sqrt{X_1^2-m_1^2F}\sqrt{X_2^2-m_2^2F}}{F}.
    \label{Ecm_radial}
\end{equation}
This expression makes the near-horizon behavior transparent.  The denominator tends to zero at the horizon, but the numerator may vanish at the same rate.  Whether $E_{\rm cm}$ is finite or divergent depends on the signs $\sigma_i$ and on the horizon values $X_{iH}=E_i-q_i\Phi_H$.

\subsection{Same-direction collisions}

If both particles are ingoing, $\sigma_1\sigma_2=+1$.  Suppose first that both are usual particles, meaning $X_{1H}>0$ and $X_{2H}>0$. Near a nonextremal horizon, $Z_i=X_i-FH_i/(2X_i)+O(F^2)$.  The numerator in Eq.~\eqref{Ecm_general} becomes
\begin{equation}
    X_1X_2-Z_1Z_2=\frac{F}{2}\left(\frac{H_1X_2}{X_1}+\frac{H_2X_1}{X_2}\right)+O(F^2).
    \label{same_expansion}
\end{equation}
Therefore the horizon limit is finite:
\begin{equation}
    E_{\rm cm,H}^2=m_1^2+m_2^2+\frac{H_{1H}X_{2H}}{X_{1H}}+\frac{H_{2H}X_{1H}}{X_{2H}}-\frac{2L_1L_2}{r_+^2}.
    \label{Ecm_usual_horizon}
\end{equation}
The apparent divergence cancels because two usual ingoing particles become asymptotically parallel as they cross the horizon. A large same-direction collision requires at least one small horizon value $X_{H}$.  For particle 1, the critical condition reads 
\begin{equation}
    X_{1H}=0
    \quad \Longleftrightarrow\quad
    E_1=q_1\Phi_H.
    \label{critical_condition}
\end{equation}
For a given energy, the critical charge is
\begin{equation}
    q_{\rm c}=\frac{E}{\Phi_H}
    =\frac{E(1-\ell)r_+}{Q}.
    \label{critical_charge}
\end{equation}
The Lorentz-violating parameter therefore changes the amount of charge required to tune a particle to criticality.  In the benchmark examples below, $\Phi_H$ increases with $\ell$, so $q_{\rm c}$ decreases.

\subsection{Head-on collisions}

If one particle is ingoing and the other is outgoing, then $\sigma_1\sigma_2=-1$.  For two usual particles,
\begin{equation}
    X_1X_2+Z_1Z_2=2X_{1H}X_{2H}+O(F),
    \label{headon_expansion}
\end{equation}
so Eq.~\eqref{Ecm_general} gives
\begin{equation}
    E_{\rm cm}^2\sim \frac{4X_{1H}X_{2H}}{F(r)}
    \qquad (r\to r_+).
    \label{headon_div}
\end{equation}
Thus a head-on collision arbitrarily close to any horizon has formally divergent center-of-mass energy.  The physical issue is how the outgoing particle is produced near the horizon.  In a classical collapse geometry there is no generic outgoing geodesic emerging from the future horizon.  It must be created by a previous scattering event, by a turning point outside the horizon, or by a white-hole-like initial condition.  This distinction is standard in discussions of super-Penrose processes and remains important here.

The critical condition \eqref{critical_condition} is only the first step.  A critical or near-critical particle must actually be able to reach the collision radius.  The answer differs between nonextremal and extremal horizons.

\subsection{Nonextremal horizons}

For a nonextremal horizon,
\begin{equation}
    F(r)=2\kappa(r-r_+)+O[(r-r_+)^2],
    \qquad
    \kappa>0.
    \label{nonext_F}
\end{equation}
For a critical particle, $X_H=0$ and
\begin{equation}
    X(r)=X'_H(r-r_+)+O[(r-r_+)^2],
    \qquad
    X'_H=-q\Phi'(r_+)=\frac{qQ}{(1-\ell)r_+^2}.
    \label{Xcrit_nonext}
\end{equation}
The radial quantity is
\begin{equation}
    Z^2=X^2-FH
    =-2\kappa H_H(r-r_+)+O[(r-r_+)^2],
    \label{Z_nonext_crit}
\end{equation}
which is negative sufficiently close to the horizon.  Therefore an exactly critical massive particle cannot reach a nonextremal horizon.  A near-critical particle can approach a turning point outside the horizon, and collisions near that turning point can be large but finite.  This is the static charged analogue of the well-known obstruction to exact critical motion near nonextremal horizons.

A near-critical particle has
\begin{equation}
    X_H=\delta,
    \qquad
    0<\delta\ll1.
    \label{nearcrit_delta}
\end{equation}
If it collides with a usual particle close to its turning point, the center-of-mass energy scales approximately as
\begin{equation}
    E_{\rm cm}^2\propto \frac{1}{\delta}
    \label{nearcrit_scaling}
\end{equation}
up to factors determined by $m_i$, $L_i$ and the local slope of $F$.  Hence nonextremal dyonic Kalb--Ramond black holes can act as strong accelerators, but the ideal divergence is replaced by a large finite value unless an outgoing particle is already present.

\subsection{Extremal horizons}

In the extremal limit $M^2=\alpha Q_{\rm eff}^2$, the outer and inner horizons coincide at
\begin{equation}
    r_e=\frac{M}{\alpha}.
    \label{re_ext}
\end{equation}
The metric function factorizes as
\begin{equation}
    F(r)=\alpha\left(1-\frac{r_e}{r}\right)^2.
    \label{F_ext_factor}
\end{equation}
Close to the horizon,
\begin{equation}
    F(r)=\frac{\alpha}{r_e^2}(r-r_e)^2+O[(r-r_e)^3].
    \label{F_ext_near}
\end{equation}
For a critical particle, $X_e=0$ and
\begin{equation}
    X(r)=\frac{qQ}{(1-\ell)r_e^2}(r-r_e)+O[(r-r_e)^2].
    \label{X_ext_near}
\end{equation}
The radial quantity becomes
\begin{equation}
    Z^2=(r-r_e)^2\left[\frac{q^2Q^2}{(1-\ell)^2r_e^4}-\frac{\alpha}{r_e^2}\left(m^2+\frac{L^2}{r_e^2}\right)\right]+O[(r-r_e)^3].
    \label{Z_ext}
\end{equation}
Thus the critical particle can reach the extremal horizon if
\begin{equation}
    \frac{q^2Q^2}{(1-\ell)^2r_e^2}\geq \alpha\left(m^2+\frac{L^2}{r_e^2}\right).
    \label{reachability_ext}
\end{equation}
Using the critical condition $E=q\Phi_H=qQ/[(1-\ell)r_e]$, this can be written as
\begin{equation}
    E^2\geq \alpha\left(m^2+\frac{L^2}{r_e^2}\right).
    \label{reachability_ext_E}
\end{equation}
When this condition holds, a collision between the critical particle and a usual particle has the BSW divergence
\begin{equation}
    E_{\rm cm}^2\propto \frac{1}{r-r_e}.
    \label{BSW_ext_scaling}
\end{equation}
The coefficient depends on the usual particle's $X_H$ and on the finite slope of the critical trajectory.  The divergence is a local test-particle result; it is expected to be softened by finite-size effects, electromagnetic radiation, self-force and backreaction.

\subsection{Collisional electric Penrose process}

The electric Penrose process may also occur through a collision rather than a decay.  Suppose particles 1 and 2 collide at radius $r_c$ and produce particles 3 and 4:
\begin{equation}
    1+2\to3+4.
    \label{reaction}
\end{equation}
Conservation laws give
\begin{align}
    E_1+E_2&=E_3+E_4,\label{coll_E_cons}\\
    q_1+q_2&=q_3+q_4,\label{coll_q_cons}\\
    g_1+g_2&=g_3+g_4,\label{coll_g_cons}\\
    L_1+L_2&=L_3+L_4.\label{coll_L_cons}
\end{align}
If particle 4 falls into the horizon with $E_4<0$, then the escaping particle can have
\begin{equation}
    E_3=E_1+E_2-E_4>E_1+E_2.
    \label{E3_extract}
\end{equation}
The collisional efficiency is
\begin{equation}
    \eta_{\rm coll}=\frac{E_3}{E_1+E_2}=1-\frac{E_4}{E_1+E_2}.
    \label{eta_coll}
\end{equation}
Near the horizon, the purely electric ideal bound is
\begin{equation}
    \eta_{\rm coll}\lesssim 1-\frac{q_4\Phi_H}{E_1+E_2},
    \qquad q_4\Phi_H<0.
    \label{eta_coll_bound}
\end{equation}
For a dyonic product, the corresponding electromagnetic bound is
\begin{equation}
    \eta_{\rm coll}^{\rm em}\lesssim 1-\frac{q_4\Phi_H+g_4\Psi_H}{E_1+E_2},
    \qquad q_4\Phi_H+g_4\Psi_H<0.
    \label{eta_coll_em_bound}
\end{equation}
This equation resembles the decay bounds \eqref{eta_bound} and \eqref{eta_em_bound}, but a collision can generate the required charge and momentum distribution dynamically.

It is tempting to identify arbitrarily large $E_{\rm cm}$ with arbitrarily large $E_3$.  This identification is generally false.  The local energy available in the center-of-mass frame is redshifted relative to infinity, and the escaping fragment must climb out through the electromagnetic and gravitational potentials.  In a same-direction BSW collision near an extremal horizon, the large local energy may be carried mainly by particles whose Killing energies remain finite.  A true super-Penrose outcome requires both high local energy and a kinematic channel that assigns sufficiently negative conserved energy to the absorbed product.  In the present static charged geometry that assignment is controlled by $q_4\Phi_H$.

\section{Numerical analysis and physical discussion}\label{sec:numerics}

In this section, we will discuss numerically the results obtained in the prior sections. In this way, to illustrate the analytic formulas we use the benchmark values $M=1$, $Q=0.30$ and $p=0.20$ which keep the black hole safely away from extremality over the range shown.  These values are also useful because they have been used in the wave-optical continuation of the same geometry. Table~\ref{tab:horizon_potential} shows how the horizon potential and the critical charge change with the LV  parameter.  The effective charge $Q_{\rm eff}^2$ grows with $\ell$, the horizon radius decreases, and the electrostatic potential $\Phi_H$ increases.  As a result, the critical charge $q_c=E/\Phi_H$ for a particle with $E=1$ decreases.

\begin{table}[t]
\caption{Horizon and electric-Penrose diagnostics for $M=1$, $Q=0.30$ and $p=0.20$.  The last two columns show the idealized near-horizon efficiencies for an infalling fragment with $q_1=-5$ and $q_1=-10$, assuming $E_0=1$ in Eq.~\eqref{eta_bound}.}
\label{tab:horizon_potential}
\begin{ruledtabular}
\begin{tabular}{c c c c c c c c}
$\ell$ & $\alpha$ & $Q_{\rm eff}^2$ & $r_+/M$ & $\Phi_H$ & $q_c(E=1)$ & $\eta_{q_1=-5}$ & $\eta_{q_1=-10}$ \\
\hline
$0.00$ & $1.0000$ & $0.1300$ & $1.9327$ & $0.1552$ & $6.4425$ & $1.7761$ & $2.5522$ \\
$0.05$ & $1.0526$ & $0.1442$ & $1.8250$ & $0.1730$ & $5.7790$ & $1.8652$ & $2.7304$ \\
$0.10$ & $1.1111$ & $0.1611$ & $1.7155$ & $0.1943$ & $5.1464$ & $1.9715$ & $2.9431$ \\
$0.15$ & $1.1765$ & $0.1817$ & $1.6037$ & $0.2201$ & $4.5438$ & $2.1004$ & $3.2008$ \\
$0.20$ & $1.2500$ & $0.2073$ & $1.4886$ & $0.2519$ & $3.9696$ & $2.2596$ & $3.5192$ \\
\end{tabular}
\end{ruledtabular}
\end{table}

\begin{figure*}[tbhp]
    \centering
    \includegraphics[width=\textwidth]{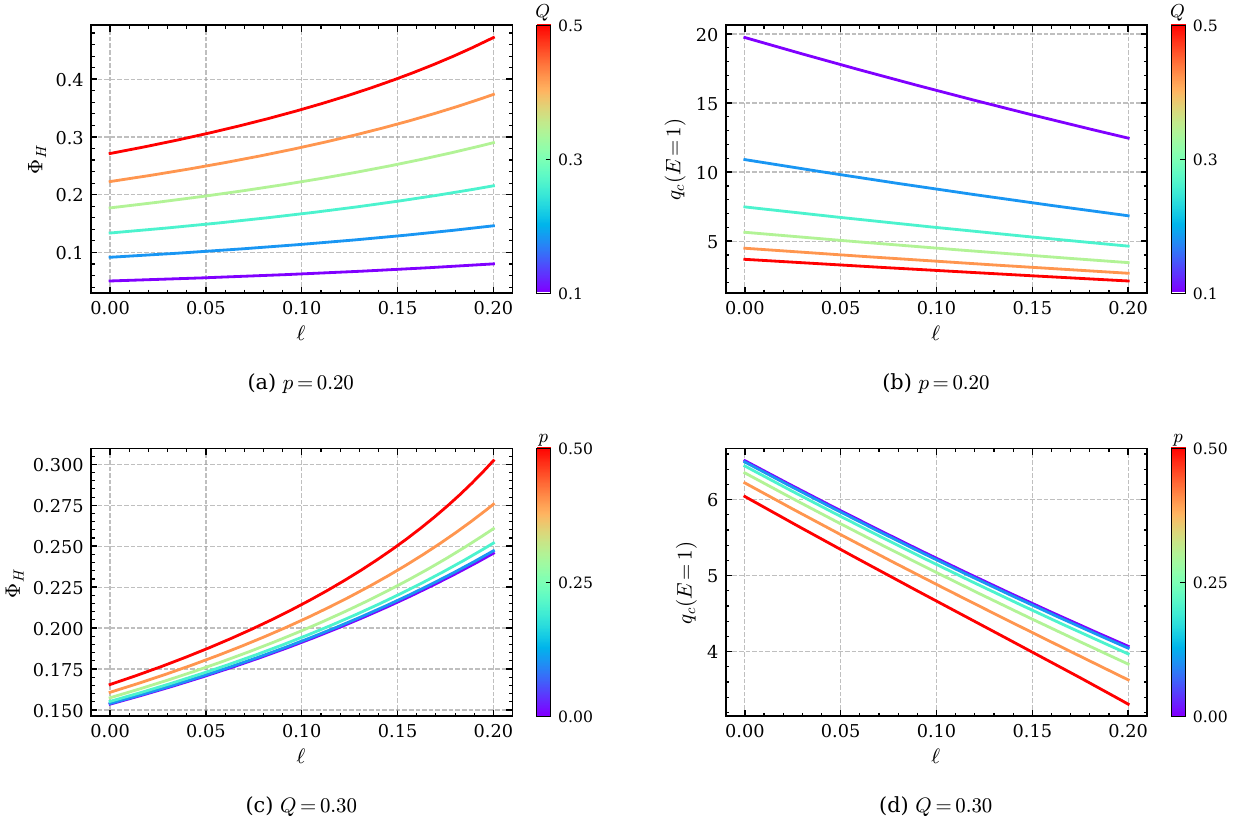}
    \caption{Horizon electrostatic potential $\Phi_H$ and critical charge $q_c=E/\Phi_H$ in the same four-panel style used in the reference article.  Panels (a) and (b) vary the electric charge $Q$ at fixed $p=0.20$, while panels (c) and (d) vary the magnetic charge $p$ at fixed $Q=0.30$.  All curves use $M=1$ and the color bars identify the swept parameter.}
    \label{fig:potential_qcrit}
\end{figure*}

\begin{figure*}[tbhp]
    \centering
    \includegraphics[width=\textwidth]{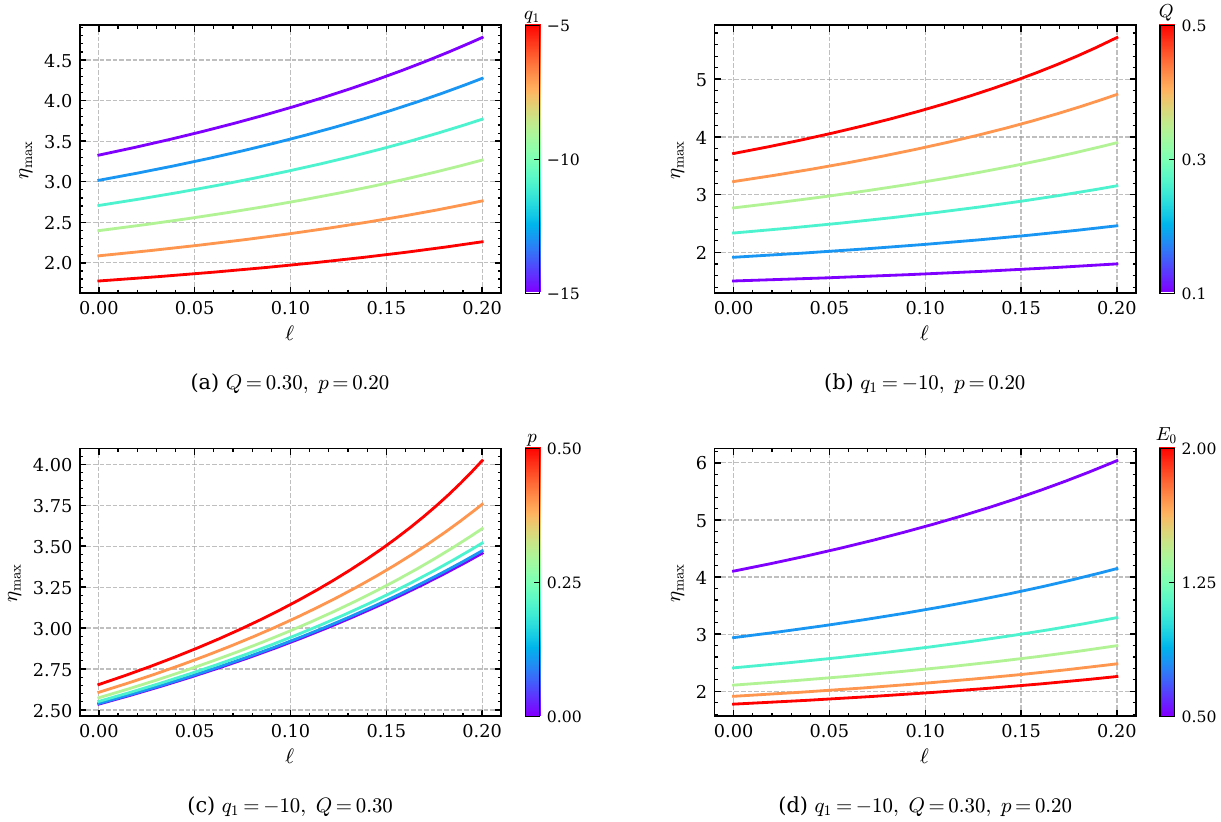}
    \caption{Ideal near-horizon electric Penrose efficiency in the reference-article plotting style.  Panel (a) varies the charge $q_1$ of the absorbed fragment, panel (b) varies the electric charge $Q$, panel (c) varies the magnetic charge $p$, and panel (d) varies the incident energy $E_0$.  The curves are kinematic upper envelopes; realistic efficiencies are reduced by momentum conservation, backreaction, self-force and escape conditions.}
    \label{fig:eta_plot}
\end{figure*}

The efficiencies in the table should not be read as realistic astrophysical predictions.  They show the direction and scale of the ideal electrostatic effect.  If one fixes the charge of the negative-energy fragment, increasing $\ell$ raises the amount of extractable energy because it increases $\Phi_H$.  Conversely, if one fixes the desired efficiency, increasing $\ell$ reduces the magnitude of charge required for the infalling negative-energy fragment.  The same trend is displayed graphically in Figs.~\ref{fig:potential_qcrit} and \ref{fig:eta_plot}.

The same benchmark also illustrates the intrinsic magnetic Penrose channel.  Table~\ref{tab:magnetic_potential} lists the magnetostatic horizon potential, the pure magnetic critical charge \(g_c=E/\Psi_H\), and two ideal near-horizon efficiencies for negative magnetic fragments.  Along the displayed sequence, \(\Psi_H\) grows faster than \(\Phi_H\), because the magnetic normalization contains \((1-2\ell)^{-1}\).  The critical magnetic charge therefore decreases as \(\ell\) increases, while the ideal magnetic efficiency increases.

\begin{table}[t]
\caption{Magnetic-Penrose diagnostics for $M=1$, $Q=0.30$ and $p=0.20$.  The last two columns show the ideal near-horizon efficiencies for absorbed magnetic fragments with $g_1=-5$ and $g_1=-10$, assuming $E_0=1$ in Eq.~\eqref{eta_mag_bound}.}
\label{tab:magnetic_potential}
\begin{ruledtabular}
\begin{tabular}{c c c c c c}
$\ell$ & $r_+/M$ & $\Psi_H$ & $g_c(E=1)$ & $\eta_{g_1=-5}$ & $\eta_{g_1=-10}$ \\
\hline
$0.00$ & $1.9327$ & $0.1035$ & $9.6637$ & $1.5174$ & $2.0348$ \\
$0.05$ & $1.8250$ & $0.1218$ & $8.2123$ & $1.6088$ & $2.2177$ \\
$0.10$ & $1.7155$ & $0.1457$ & $6.8619$ & $1.7287$ & $2.4573$ \\
$0.15$ & $1.6037$ & $0.1782$ & $5.6129$ & $1.8908$ & $2.7816$ \\
$0.20$ & $1.4886$ & $0.2239$ & $4.4658$ & $2.1196$ & $3.2392$ \\
\end{tabular}
\end{ruledtabular}
\end{table}

\begin{figure*}[tbhp]
    \centering
    \includegraphics[width=\textwidth]{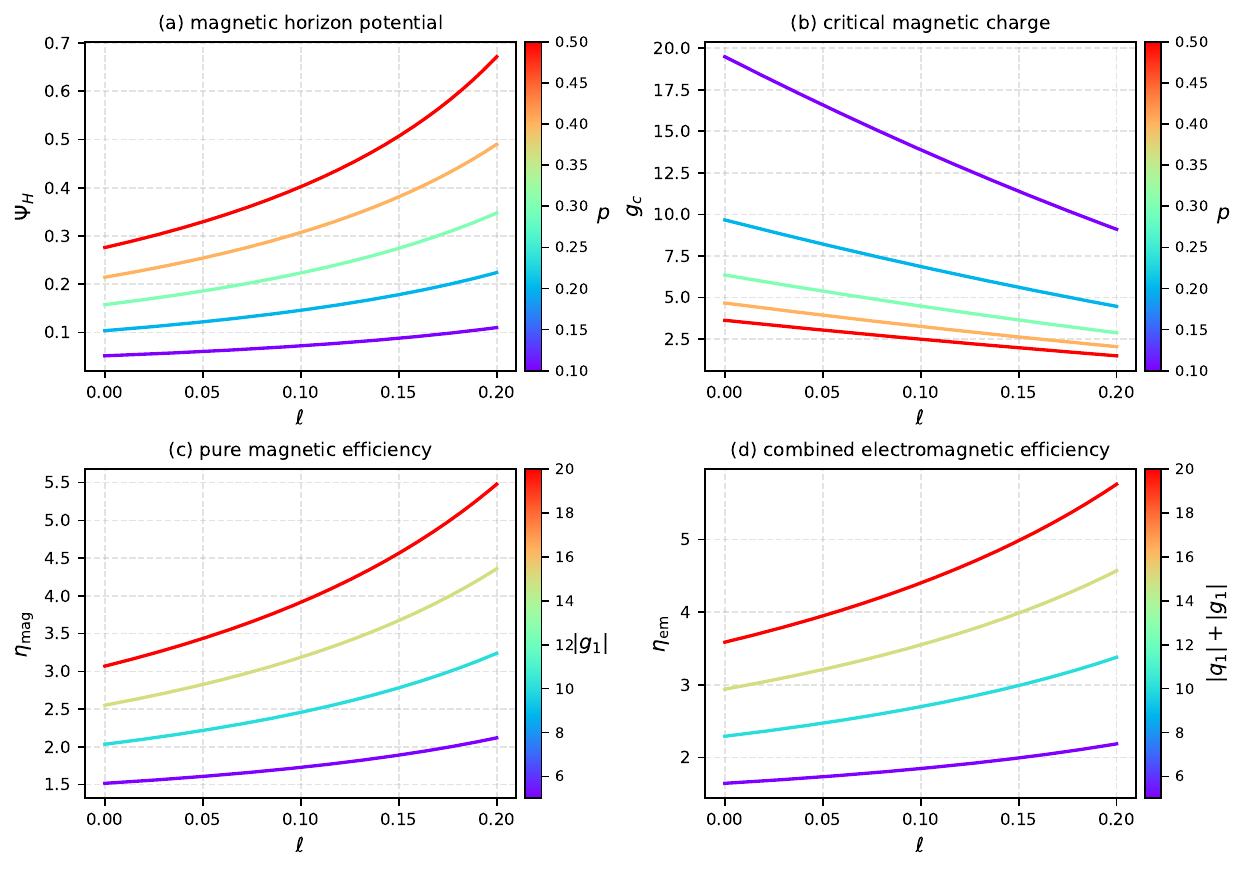}
    \caption{Magnetic Penrose diagnostics in the same four-panel style used for the other plots.  Panels (a) and (b) show the magnetostatic horizon potential \(\Psi_H\) and the critical magnetic charge \(g_c=E/\Psi_H\).  Panel (c) shows the ideal pure magnetic efficiency, while panel (d) shows the combined electromagnetic efficiency when both electric and magnetic charges are assigned to the absorbed fragment.}
    \label{fig:mag_efficiency}
\end{figure*}

\begin{figure*}[tbhp]
    \centering
    \includegraphics[width=\textwidth]{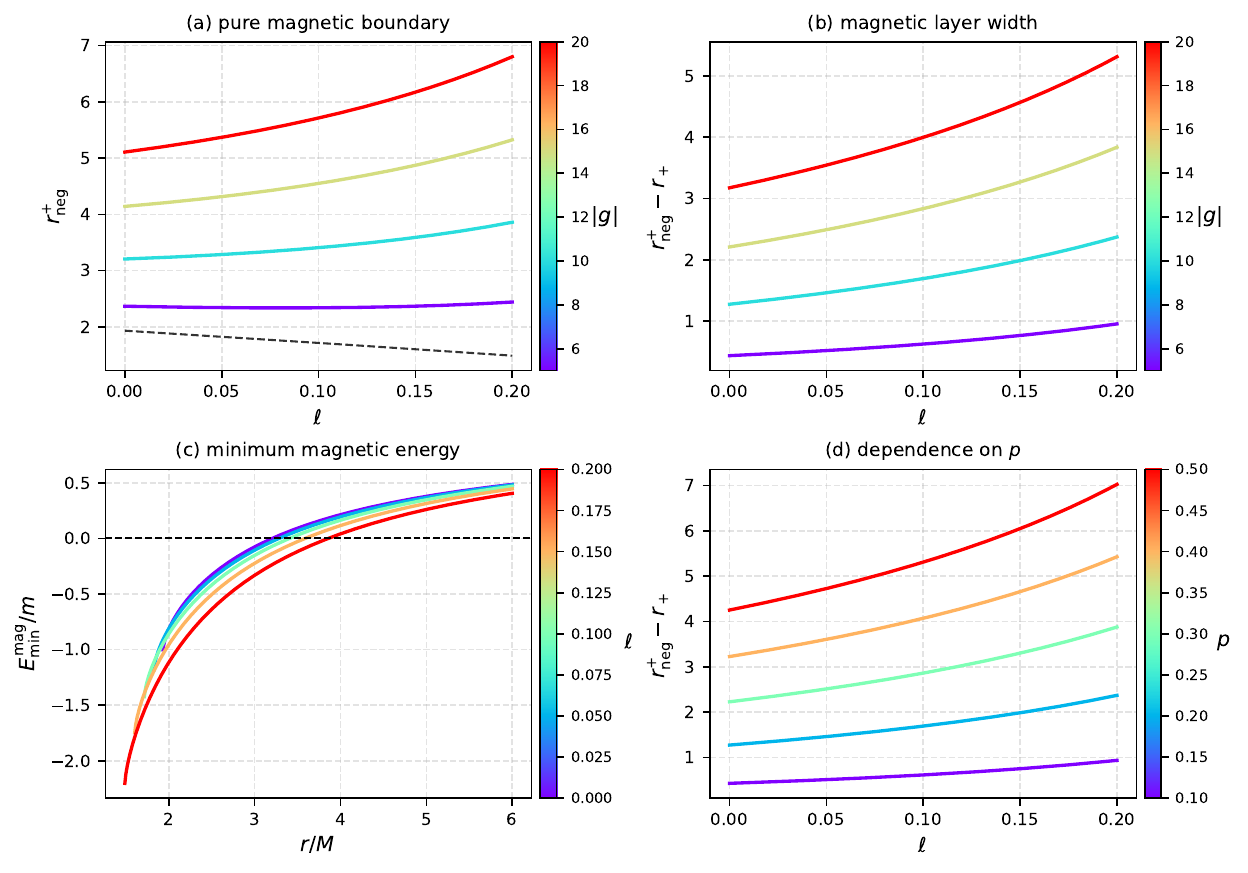}
    \caption{Negative-energy domains for the magnetic Penrose channel.  Panels (a) and (b) show the outer magnetic boundary and the corresponding layer width.  Panel (c) displays the minimum pure magnetic energy for an absorbed fragment with $g=-10$.  Panel (d) shows how the layer width changes when the black-hole magnetic charge is varied.}
    \label{fig:mag_negative_domain}
\end{figure*}

\subsection{Phase-space diagrams}

\begin{figure*}[tbhp]
    \centering
    \includegraphics[width=\textwidth]{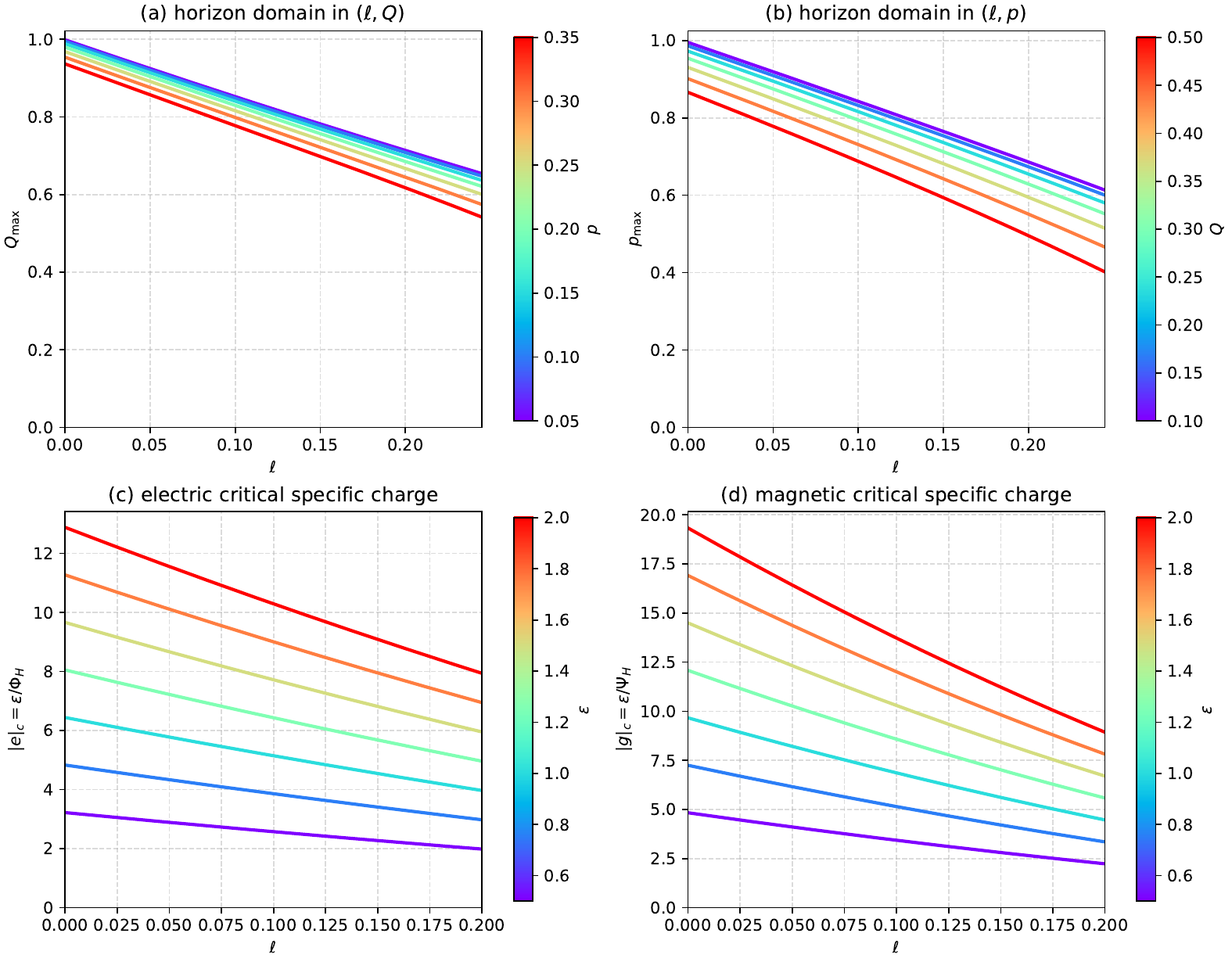}
    \caption{Phase-space boundaries associated with the black-hole existence condition and with critical probe charges.  Panels (a) and (b) show the horizon-allowed regions in the $(\ell,Q)$ and $(\ell,p)$ planes by plotting the upper envelopes $Q_{\max}(\ell,p)$ and $p_{\max}(\ell,Q)$.  Panels (c) and (d) display the critical specific electric and magnetic charges, $|e|_{c}=\varepsilon/\Phi_H$ and $|g|_{c}=\varepsilon/\Psi_H$, for the benchmark background $M=1$, $Q=0.30$ and $p=0.20$.  The color bars indicate the swept parameter in each panel.}
    \label{fig:phase_background}
\end{figure*}

It is also useful to reorganize the same information in terms of phase-space boundaries instead of radius profiles.  The first level is the background parameter space itself.  Starting from the black-hole condition
\begin{equation}
M^2\geq \alpha Q_{\rm eff}^2,
\label{bh_bound_phase}
\end{equation}
one may solve for the largest admissible electric or magnetic charge at fixed $M$:
\begin{align}
Q^2&\leq Q_{\max}^2(\ell,p)
=(1-\ell)^2\left[(1-\ell)M^2-\frac{p^2}{1-2\ell}\right],
\label{Qmax_phase}\\
 p^2&\leq p_{\max}^2(\ell,Q)
=(1-2\ell)\left[(1-\ell)M^2-\frac{Q^2}{(1-\ell)^2}\right].
\label{pmax_phase}
\end{align}
These expressions define the outer boundaries of the allowed black-hole region in the $(\ell,Q)$ and $(\ell,p)$ planes.  Figure~\ref{fig:phase_background} shows that increasing one gauge charge reduces the maximum value of the other, while increasing $\ell$ shrinks the dyonic domain because the effective charge entering the metric grows.

For the probes, the natural phase-space variables are the specific energy $\varepsilon=E/m$ and the specific electric and magnetic charges $|e|=|q|/m$ and $|g|=|g_m|/m$.  The near-horizon threshold for a negative-energy electric fragment is
\begin{equation}
\varepsilon<|e|\Phi_H,
\qquad
equivalently \qquad
|e|>|e|_{c}=\frac{\varepsilon}{\Phi_H},
\label{electric_phase_boundary}
\end{equation}
while the magnetic and combined dyonic conditions are
\begin{align}
\varepsilon&<|g|\Psi_H,
\qquad
|g|>|g|_{c}=\frac{\varepsilon}{\Psi_H},
\label{magnetic_phase_boundary}\\
\varepsilon&<|e|\Phi_H+|g|\Psi_H.
\label{dyonic_phase_boundary}
\end{align}
Thus, in the $(|e|,\varepsilon)$ and $(|g|,\varepsilon)$ planes the extraction region lies below the corresponding straight lines, whereas in the dyonic $(|e|,|g|)$ plane it lies above the boundary set by Eq.~\eqref{dyonic_phase_boundary}.  This representation makes the competition between electric and magnetic work terms especially transparent: at fixed $\varepsilon$, increasing $\ell$ lowers the critical charges and therefore enlarges the extraction domain.

\begin{figure*}[tbhp]
    \centering
    \includegraphics[width=\textwidth]{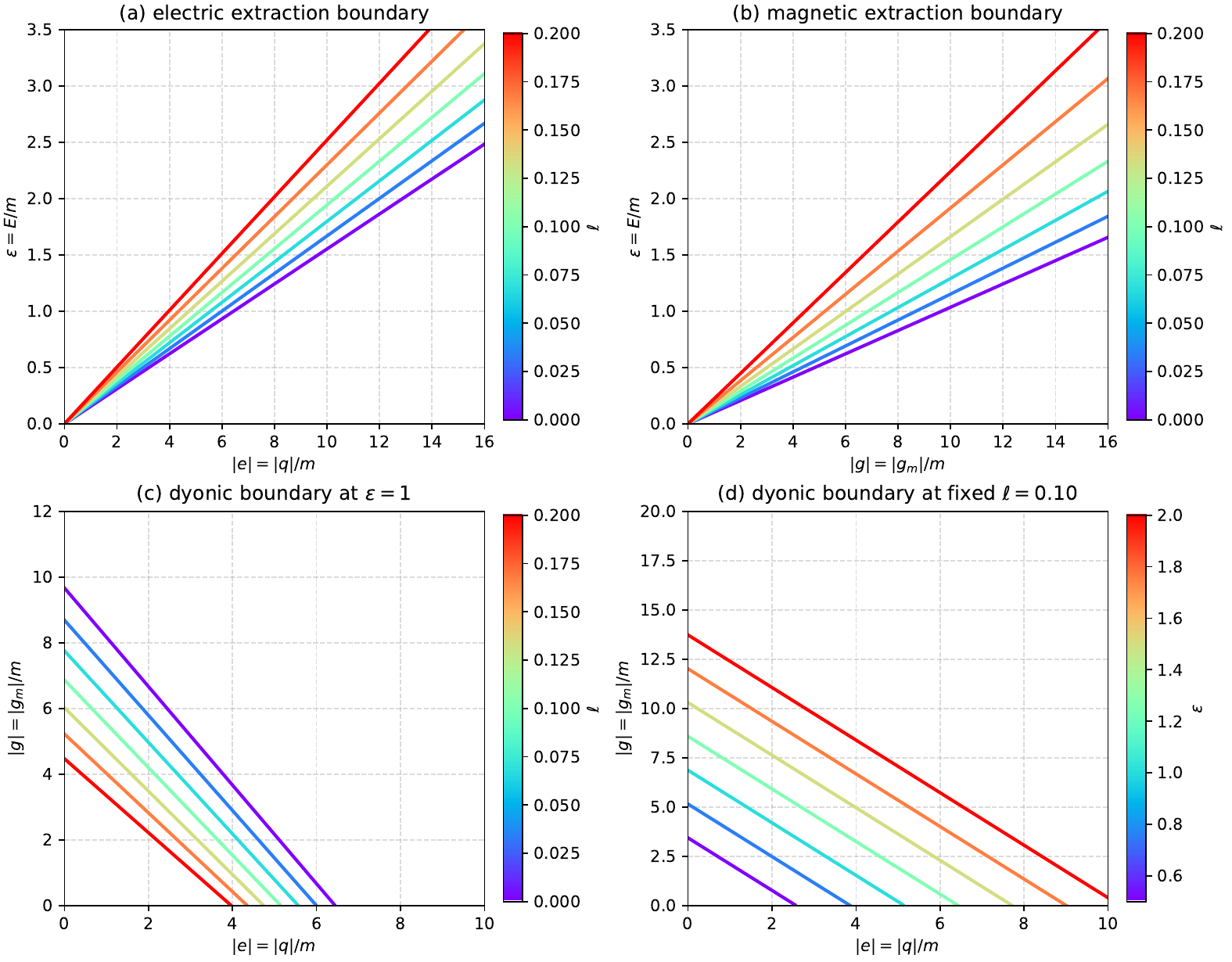}
    \caption{Probe phase-space diagrams for negative-energy extraction.  Panels (a) and (b) show the electric and magnetic boundaries in the $(|e|,\varepsilon)$ and $(|g|,\varepsilon)$ planes; the region below each line corresponds to admissible negative-energy fragments at the horizon.  Panel (c) shows the dyonic boundary in the $(|e|,|g|)$ plane at fixed $\varepsilon=1$, for which the extraction region lies above the curves.  Panel (d) shows the same dyonic boundary at fixed $\ell=0.10$ and varying specific energy.}
    \label{fig:phase_probes}
\end{figure*}

Figures~\ref{fig:phase_background} and \ref{fig:phase_probes} complement the earlier radius-based analysis in a convenient way.  In particular, panels (c) and (d) of Fig.~\ref{fig:phase_background} make it clear that the electric and magnetic critical lines behave monotonically with $\ell$ along the benchmark trajectory, while Fig.~\ref{fig:phase_probes}(c,d) shows directly how a mixed electric-magnetic assignment can compensate a smaller value of either charge.  This is useful when constructing decay or collision channels numerically, because the figures immediately identify the part of probe-parameter space in which negative-energy states are kinematically available before the full radial motion is integrated.

A second analysis concerns the criticality condition for collisions.  For $E=1$ and $Q>0$, a positive critical charge is required.  The critical particle has $X_H=0$ and is therefore the one that can generate a BSW-type enhancement in the extremal limit.  The fact that $q_c$ decreases with $\ell$ means that Lorentz violation makes the critical tuning less demanding for the chosen parameter family.  This is not universal for all possible paths through parameter space, because $r_+$ and $Q_{\rm eff}^2$ can vary in competing ways.  The robust statement is Eq.~\eqref{critical_charge}: the tuning is governed by the inverse of the horizon potential. For nonextremal black holes in Table~\ref{tab:horizon_potential}, an exactly critical massive particle cannot reach the horizon.  A near-critical particle can reach a turning point outside the horizon, and a collision with a usual ingoing particle there can be highly energetic but finite.  In numerical studies this is the regime one should use to avoid confusing formal extremal divergences with physically reachable nonextremal trajectories.

The formulas above are analytic, but they are also intended to guide direct numerical work.  A useful implementation begins with the choice of the background parameters $(M,Q,p,\ell)$ and first checks the black-hole inequality $M^2\geq \alpha Q_{\rm eff}^2$.  One then computes $r_+$, fixes the gauges by imposing $\Phi(\infty)=\Psi(\infty)=0$, and evaluates the horizon potentials $\Phi_H$ and $\Psi_H$.  For an electrically charged particle one specifies $(m,q,E,L)$ and checks $X(r)=E-q\Phi(r)>0$ outside the horizon.  For a dyonic probe one instead specifies $(m,q,g,E,L)$ and checks $X_{\rm em}(r)=E-q\Phi(r)-g\Psi(r)>0$.  The radial motion is then obtained from
\begin{equation}
R(r)\equiv X^2(r)-F(r)\left(m^2+\frac{L^2}{r^2}\right),
\label{R_numeric}
\end{equation}
with
\begin{equation}
\dot r=\sigma\frac{\sqrt{R(r)}}{m}    
\end{equation}
The physically allowed interval is the region in which $R(r)\geq0$ and $X(r)>0$.  This simple two-condition check is important because a particle that formally satisfies the critical relation $E=q\Phi_H$ may still be unable to reach a nonextremal horizon.  Numerically, this obstruction appears as a turning point outside $r_+$. For an electric Penrose decay at $r_s$, a practical algorithm is the following.  First choose the parent particle and integrate it to the splitting radius.  Second choose the charge and angular momentum assignment of the fragments so that $q_0=q_1+q_2$ and $L_0=L_1+L_2$.  Third impose the local energy bounds
\begin{equation}
    E_i\geq q_i\Phi(r_s)+\sqrt{F(r_s)}\sqrt{m_i^2+\frac{L_i^2}{r_s^2}},
    \qquad i=1,2,
\end{equation}
for fragments that are future directed at the splitting point.  Finally check radial momentum conservation using
\begin{equation}
    \sigma_0 Z_0=\sigma_1 Z_1+\sigma_2 Z_2,
    \label{radial_momentum_decay}
\end{equation}
where all $Z_i$ are evaluated at $r_s$.  The escaping fragment must additionally have an allowed radial path to infinity.  Since $\Phi(r)\to0$ and $F(r)\to\alpha$, a massive particle that reaches infinity with nonzero velocity must satisfy $E_2^2\geq \alpha m_2^2$ for radial motion.  If $E_2^2<\alpha m_2^2$, the fragment is bound even if it is locally outgoing immediately after the decay.

The negative-energy interval can be estimated before integrating any trajectory.  For radial motion with $m=1$ and $qQ<0$, the outer boundary $r_{\rm neg}^+$ is obtained from Appendix C.  Table~\ref{tab:negative_region} gives the width of this interval for the same background as Table~\ref{tab:horizon_potential}.  The width grows with $\ell$ for the displayed charge assignments.  This happens because the increase in the electrostatic term dominates the change in the gravitational barrier along this parameter path.

\begin{table}[t]
\caption{Outer boundary of the radial negative-energy region for $M=1$, $Q=0.30$, $p=0.20$, $m=1$ and $L=0$.  The interval is $r_+<r<r_{\rm neg}^+$ for an oppositely charged fragment.}
\label{tab:negative_region}
\begin{ruledtabular}
\begin{tabular}{c c c c c}
$\ell$ & $r_+/M$ & $r_{\rm neg}^+(q=-5)/M$ & $r_{\rm neg}^+(q=-10)/M$ & $r_{\rm neg}^+(q=-10)-r_+$ \\
\hline
$0.00$ & $1.9327$ & $2.7664$ & $4.1417$ & $2.2089$ \\
$0.05$ & $1.8250$ & $2.7203$ & $4.1499$ & $2.3249$ \\
$0.10$ & $1.7155$ & $2.6790$ & $4.1657$ & $2.4503$ \\
$0.15$ & $1.6037$ & $2.6431$ & $4.1901$ & $2.5864$ \\
$0.20$ & $1.4886$ & $2.6129$ & $4.2241$ & $2.7355$ \\
\end{tabular}
\end{ruledtabular}
\end{table}

\begin{figure*}[tbhp]
    \centering
    \includegraphics[width=\textwidth]{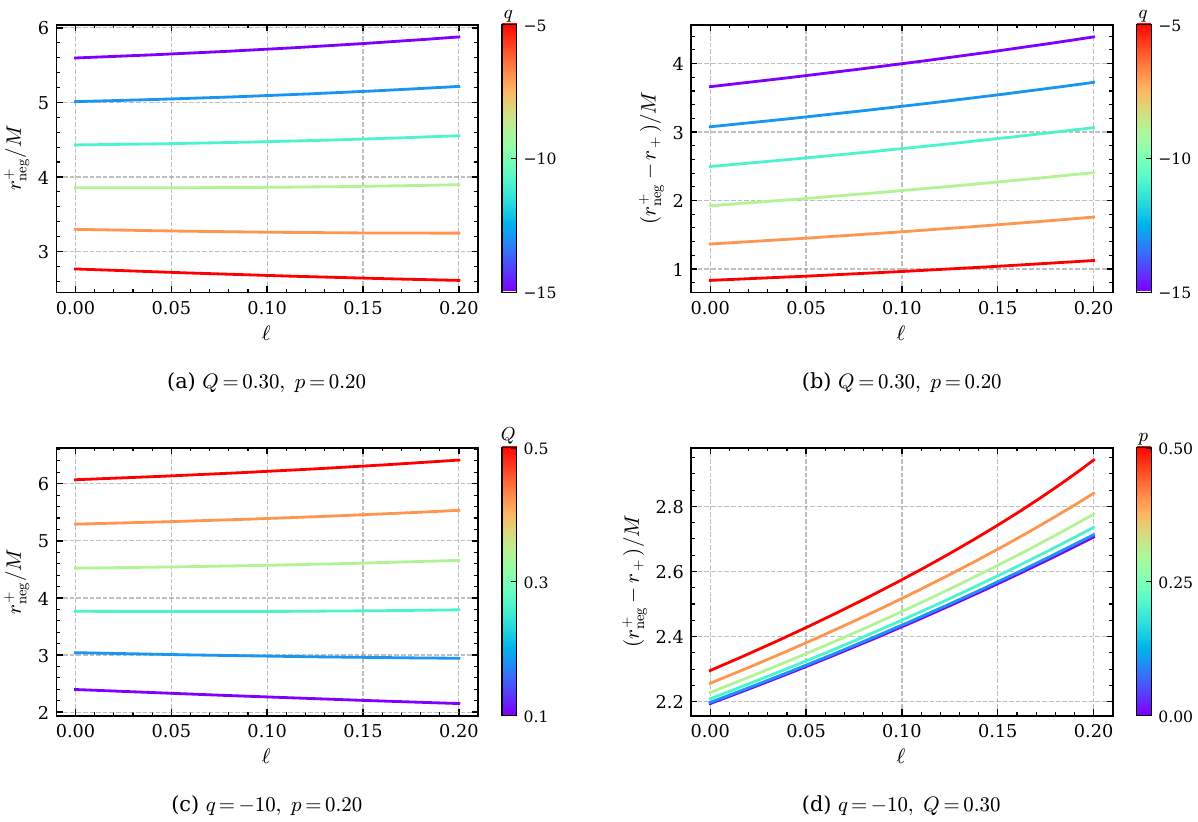}
    \caption{Radial negative-energy domain in the reference-article plotting style.  Panels (a) and (c) show the outer boundary $r_{\rm neg}^{+}$, while panels (b) and (d) show the width $r_{\rm neg}^{+}-r_+$.  The color bars display the swept particle or black-hole parameter, and the interval is physically available for $r_+<r<r_{\rm neg}^{+}$.}
    \label{fig:neg_domain}
\end{figure*}

\begin{figure*}[tbhp]
    \centering
    \includegraphics[width=\textwidth]{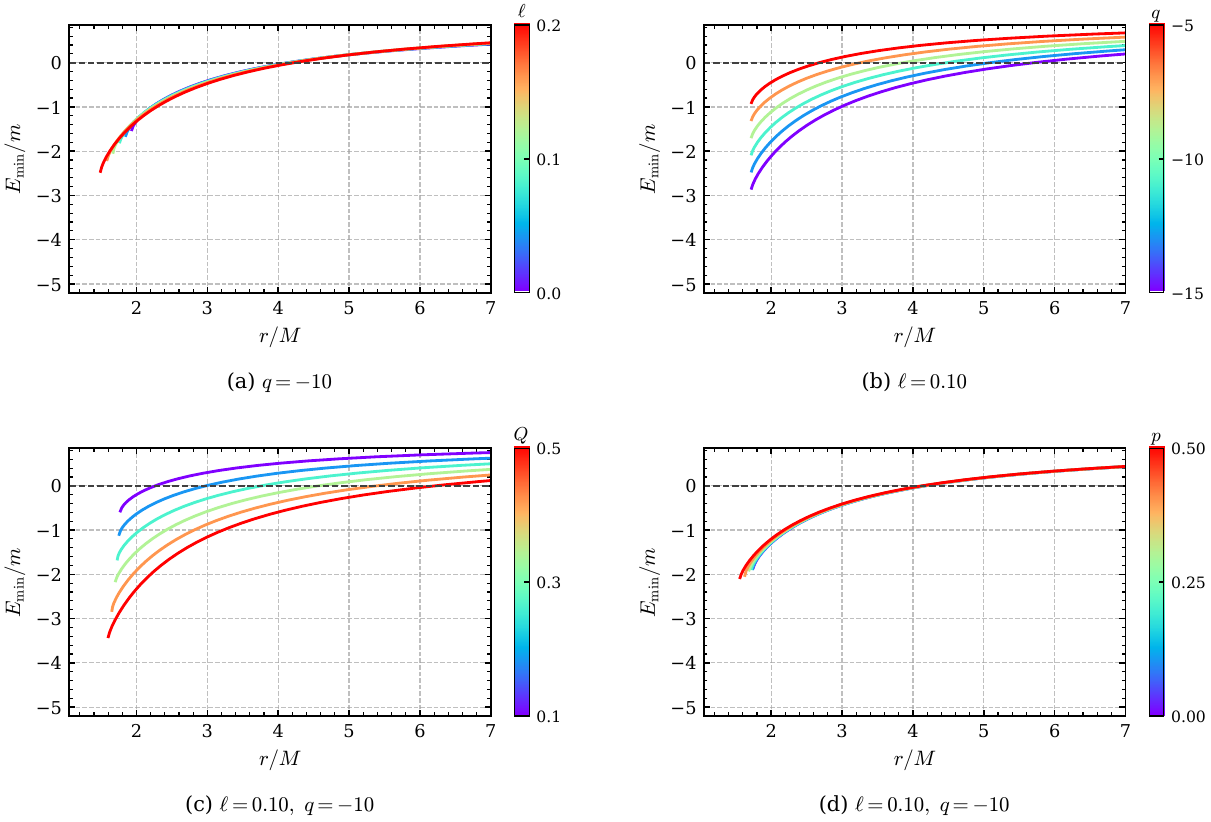}
    \caption{Minimum radial energy per unit mass for oppositely charged fragments.  The dashed horizontal line marks $E_{\min}=0$; curves below it identify radii where negative-energy states are locally available.  The panels vary, respectively, $\ell$, the fragment charge $q$, the electric charge $Q$, and the magnetic charge $p$.}
    \label{fig:emin_profiles}
\end{figure*}

Figures~\ref{fig:neg_domain} and \ref{fig:emin_profiles} show the same information in a more diagnostic form.  The horizon radius decreases along the benchmark sequence, while the outer boundary of the negative-energy interval either stays nearly fixed or moves outward, depending on the fragment charge.  The result is a wider negative-energy layer for larger $\ell$.

\begin{figure*}[tbhp]
    \centering
    \includegraphics[width=\textwidth]{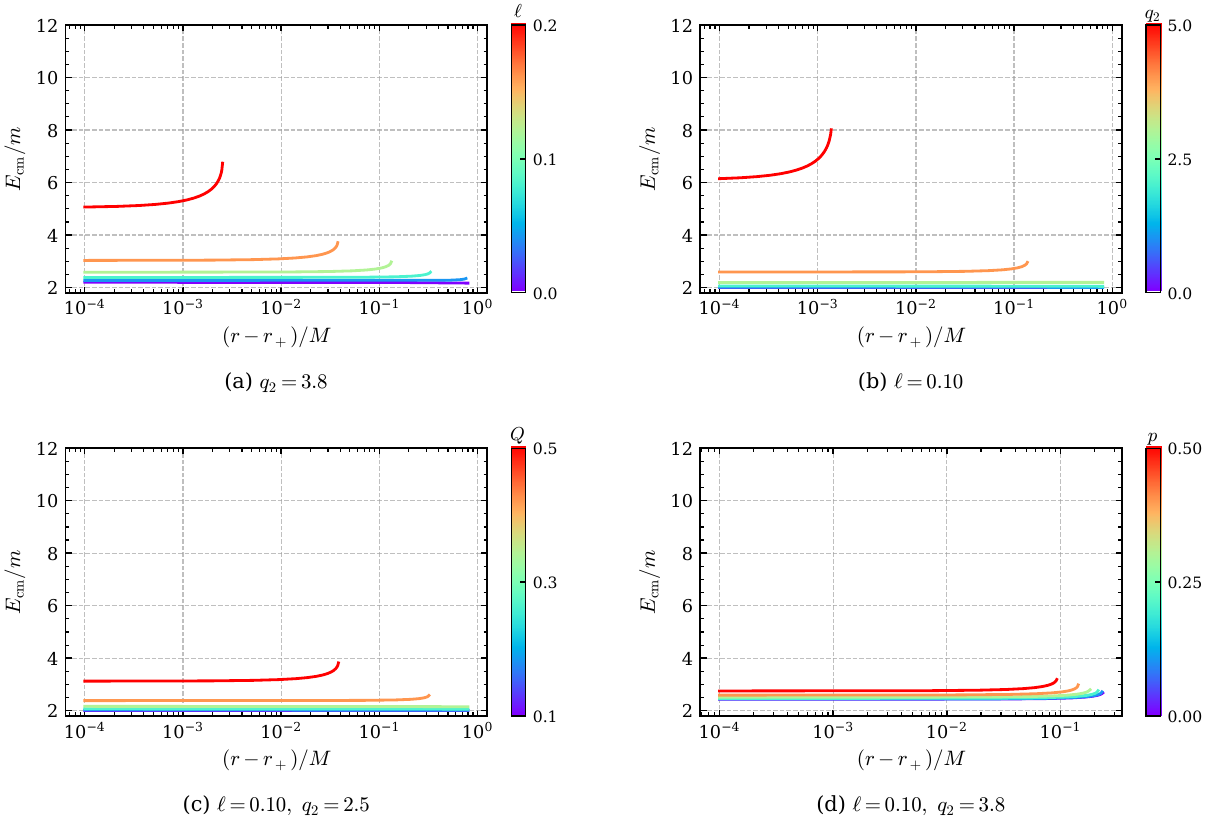}
    \caption{Center-of-mass energy for same-direction radial collisions in the reference-article plotting style.  A neutral usual particle $(E_1,m_1,q_1)=(1,1,0)$ collides with a charged particle $(E_2,m_2)=(1,1)$.  The panels vary $\ell$, $q_2$, $Q$, and $p$, respectively; dynamically forbidden points with $Z_i^2<0$ are omitted.}
    \label{fig:ecm_plot}
\end{figure*}

For collision calculations the same admissibility logic applies.  The expression for $E_{\rm cm}$ is local, but both incoming trajectories must be dynamically allowed at the collision radius.  If one wishes to model a collisional Penrose event, the product assigned to fall into the black hole should satisfy the negative-energy condition and should have inward radial momentum at the event.  The product assigned to escape must have no outer turning point between $r_c$ and infinity.  These requirements are sometimes more restrictive than the algebraic conservation laws, and they are the main reason why very large center-of-mass energy does not automatically translate into very large energy at infinity.  Figure~\ref{fig:ecm_plot} illustrates this point for a same-direction radial collision: the enhancement grows as the charged particle approaches the critical charge, but the plotted branches are restricted to radii where the radial motion is admissible.

A final numerical point concerns units at infinity.  The coordinate $t$ is the natural time variable of the exact solution, but an observer who wants the asymptotic metric coefficient of the time coordinate to be $-1$ may introduce $T=\sqrt{\alpha}\,t$.  If the canonical momentum conjugate to $t$ is $E$, the energy conjugate to $T$ is
\begin{equation}
    E_T=\frac{E}{\sqrt{\alpha}},
    \qquad
    \Phi_T(r)=\frac{\Phi(r)}{\sqrt{\alpha}}.
    \label{normalized_energy}
\end{equation}
The criticality condition is invariant under this rescaling:
\begin{equation}
    E-q\Phi_H=0
    \quad\Longleftrightarrow\quad
    E_T-q\Phi_{T,H}=0.
\end{equation}
The efficiency ratios used above are also invariant, provided the same clock is used for both the initial and final energies.  In contrast, absolute statements such as the threshold for a particle to be unbound at infinity depend on whether one writes them in the $t$ or $T$ convention.  In the $t$ convention a massive radial particle reaching infinity must satisfy $E^2\geq \alpha m^2$, whereas in the normalized convention the same condition is simply $E_T^2\geq m^2$.

The electromagnetic gauge choice is independent of this time normalization.  We keep $A_t=-\Phi(r)$ with $\Phi(\infty)=0$, so that the conserved energy is measured relative to a detector at infinity.  A constant gauge shift $\Phi\to\Phi+C$ would transform the canonical energy as $E\to E+qC$, leaving $X=E-q\Phi$ unchanged.  The local trajectory would therefore be the same, but the interpretation of negative energy relative to infinity would be obscured unless the boundary value of the potential were restored.  For this reason all extraction formulae should be written in terms of the physical potential difference between the horizon and infinity, which in the present gauge is exactly $\Phi_H$. This convention is especially useful when comparing different values of $\ell$.  Since $\alpha$ changes with $\ell$, one might otherwise mistake a change of units for a change of physics.  The coordinate quantities in Tables I and II are internally consistent because they use the same convention as the seed metric.  To convert the tabulated potential to the normalized-time convention, divide every value of $\Phi_H$ by $\sqrt{\alpha}$.  The critical charge $q_c=E/\Phi_H$, however, is unchanged if the particle energy is converted at the same time.

The electric Penrose process is sensitive to the horizon potential $\Phi_H$, while neutral wave observables such as the eikonal ringdown and the shadow are mostly sensitive to $\alpha$ and $Q_{\rm eff}^2$.  Combining particle and wave diagnostics can therefore help break degeneracies, but it does not remove them completely. The metric sees the electric and magnetic charges through
\begin{equation}
    Q_{\rm eff}^2=\frac{Q^2}{(1-\ell)^2}+\frac{p^2}{1-2\ell},
    \label{Qeff_degeneracy}
\end{equation}
whereas the electric Penrose process sees the electrostatic potential
\begin{equation}
    \Phi_H=\frac{Q}{(1-\ell)r_+}.
    \label{Phi_degeneracy}
\end{equation}
The magnetic charge affects $\Phi_H$ only through $r_+$.  Thus a measurement of an electric extraction channel would be more directly sensitive to $Q$ than a shadow measurement, while a shadow or high-frequency absorption measurement would be sensitive to the combination $Q_{\rm eff}^2$.  If both types of information were available, one could in principle separate $Q$ from $p$ more efficiently than with either channel alone.

There is also a degeneracy involving the normalization of time.  Since $F(\infty)=\alpha$, the coordinate energy $E$ used throughout the paper differs from the energy measured with the normalized asymptotic time $T=\sqrt{\alpha}t$.  For dimensionless ratios such as $q\Phi_H/E$ this distinction cancels if all quantities are transformed consistently.  For comparing absolute frequencies, temperatures or energies with observations, however, the normalization must be specified. Another important issue is gauge.  The negative-energy condition is meaningful only after fixing the electrostatic potential at infinity.  We have used $\Phi(\infty)=0$, which is the natural gauge for scattering and for defining energy extraction relative to infinity.  In thermodynamic discussions one often uses potential differences between the horizon and infinity; our $\Phi_H$ is precisely that difference.

Finally, the test-particle approximation must not be overinterpreted.  Very large charge-to-mass ratios, very large local center-of-mass energies and arbitrarily close near-horizon splittings are idealizations.  Self-force, pair production, discharge of the black hole, finite-size constraints and backreaction can all reduce the practical efficiency.  The value of the calculation is that it identifies the clean kinematic channels and the precise way in which $\ell$, $Q$ and $p$ enter them.

Let us check some limits. First, setting $\ell=0$ gives
\begin{equation}
    F(r)=1-\frac{2M}{r}+\frac{Q^2+p^2}{r^2},
    \qquad
    \Phi(r)=\frac{Q}{r}.
    \label{RN_limit}
\end{equation}
All expressions reduce to those of the ordinary dyonic Reissner-Nordstrom black hole.  The horizon potential becomes $\Phi_H=Q/r_+$ and the critical charge is $q_c=Er_+/Q$.  The magnetic charge contributes to the horizon position but not to the electrostatic work. If $Q=p=0$ while $\ell\neq0$, then
\begin{equation}
    F(r)=\alpha-\frac{2M}{r},
    \qquad
    r_+=\frac{2M}{\alpha},
    \qquad
    \Phi_H=0.
    \label{neutral_KR_limit}
\end{equation}
There is no electric Penrose process for electrically charged particles because there is no electric potential.  Particle collisions still occur, but there is no electromagnetic negative-energy channel.  This limit is a useful reminder that Lorentz violation alone does not produce negative canonical energy in a static spacetime. If $p=0$, the magnetic term disappears from $Q_{\rm eff}^2$:
\begin{equation}
    Q_{\rm eff}^2=\frac{Q^2}{(1-\ell)^2}.
    \label{pure_electric_limit}
\end{equation}
The electric Penrose process remains active and is controlled by $\Phi_H=Q/[(1-\ell)r_+]$.  This is the cleanest setting for isolating the LV correction to charged-particle extraction. If $Q=0$ but $p\neq0$, then
\begin{equation}
    \Phi(r)=0,
    \qquad
    Q_{\rm eff}^2=\frac{p^2}{1-2\ell}.
    \label{pure_magnetic_limit}
\end{equation}
The metric is charged, but the electric Penrose process is absent for electrically charged particles because $E=mF\dot t$ is nonnegative outside the horizon.  The magnetic Penrose process remains active if magnetically charged probes exist, and it is controlled by $\Psi_H=p/[(1-2\ell)r_+]$.  Thus the pure magnetic black hole is a clean limit in which the electric and magnetic extraction channels are completely separated.

\section{Charged scalar sector and superradiant scattering}
\label{sec:charged}

As a next step, we will in this section develop the wave-extraction part of the analysis by studying the effect of superradiance.  The electric Penrose process discussed above is formulated in terms of charged massive particles, but the same physical ingredient appears in wave scattering: the conserved flux associated with the stationary Killing vector can become negative at the horizon.  This is the charged-field version of superradiance.  In the present static spacetime there is no rotational term \(\omega-m\Omega_H\), so the only possible work term is electrostatic.  The magnetic charge affects the result indirectly through the geometry and directly through the angular monopole spectrum, but it does not by itself create an electric superradiant window for an electrically charged scalar.

As a starting point, let us consider a massless complex scalar field with charge \(q_s\).  It obeys the following equation
\begin{equation}
 \left(D_\mu D^\mu\right)\Psi=0,
\label{chargedKG}
\end{equation}
with $D_\mu=\nabla_\mu-iq_s A_\mu$. We keep the same gauge used in the particle analysis, i.e.
\begin{equation}
 A_\mu dx^\mu=-\Phi(r)dt+p\cos\theta\,d\phi,
 \label{gaugepotential}
\end{equation}
with
\begin{equation}
    \Phi(r)=\frac{Q}{(1-\ell)r}
\end{equation}
so that \(\Phi(\infty)=0\) and \(\Phi_H=\Phi(r_+)\).  This gauge choice is not a cosmetic detail.  The sign of the horizon flux is controlled by the potential difference between infinity and the horizon, and a constant shift of \(A_t\) would change the canonical frequency labels unless the boundary condition at infinity were imposed again. The angular part is not described by ordinary spherical harmonics when \(p\neq0\).  Defining $s=q_s p$, the angular functions are monopole harmonics \(Y^{(s)}_{jm}(\theta,\phi)\) \cite{WuYang1976,Kazama1977}, with
\begin{equation}
 j=|s|, |s|+1, |s|+2,\ldots,
 \qquad
 m=-j,-j+1,\ldots,j .
\end{equation}
Let us adopt the following separation
\begin{equation}
 \Psi=\frac{1}{r}\psi_j(r)Y^{(s)}_{jm}(\theta,\phi)e^{-i\omega t},
\end{equation}
which gives the radial equation, namely
\begin{equation}
 \frac{d^2\psi_j}{dr_*^2}+\left[(\omega-q_s\Phi)^2-V^{(q)}_j(r)\right]\psi_j=0,
\label{chargedradial}
\end{equation}
where $\frac{dr_*}{dr}=\frac{1}{F(r)}$ and
\begin{equation}
 V^{(q)}_j(r)=F(r)\left[\frac{j(j+1)-s^2}{r^2}+\frac{F'(r)}{r}\right].
 \label{chargedpotential}
\end{equation}
Thus, the parameter \(p\) enters in two ways. Firstly, it contributes to \(Q_{\rm eff}^2\) in the metric and secondly it shifts the allowed angular eigenvalues through \(s=q_sp\).  The electric charge \(Q\), however, is the parameter that appears in the horizon work term \(q_s\Phi_H\).  If Eq.~\eqref{chargedradial} is written as follows
\begin{equation}
 \frac{d^2\psi_j}{dr_*^2}+\left[\omega^2-\mathcal U_{\omega j}(r)\right]\psi_j=0,
\end{equation}
then the frequency-dependent barrier is
\begin{equation}
 \mathcal U_{\omega j}(r)=V^{(q)}_j(r)+2\omega q_s\Phi(r)-q_s^2\Phi^2(r).
 \label{frequencybarrier}
\end{equation}
This form is useful for visualization, but it should not be confused with an eigenvalue-independent scattering potential.  Its dependence on \(\omega\) simply reflects the minimal coupling between the scalar field and the electrostatic background. Near the event horizon, the potential term proportional to \(F(r)\) vanishes and the radial solution behaves as
\begin{equation}
 \psi_j\sim {\cal T}\,e^{-i(\omega-q_s\Phi_H)r_*},
 \qquad r\rightarrow r_+,
 \label{horizonwave}
\end{equation}
where the sign is chosen to impose a purely ingoing wave at the future horizon.  At infinity one has \(\Phi\to0\) and \(F\to\alpha\), so the scattering solution is
\begin{equation}
 \psi_j\sim {\cal I}e^{-i\omega r_*}+{\cal R}e^{+i\omega r_*},
 \qquad r\rightarrow \infty .
 \label{infinitywave}
\end{equation}
The conical normalization changes the interpretation of the asymptotic clock, but it does not change the flux balance if the same time coordinate is used consistently.  The conserved Wronskian of Eq.~\eqref{chargedradial} gives
\begin{equation}
 |{\cal R}|^2=|{\cal I}|^2-
 \frac{\omega-q_s\Phi_H}{\omega}|{\cal T}|^2 .
 \label{wronskian}
\end{equation}
This relation is the charged analogue of the standard superradiance formula \cite{PressTeukolsky1972,Bekenstein1973,FuruhashiNambu2004}.  The reflected amplitude is larger than the incident amplitude when
\begin{equation}
 \omega\left(\omega-q_s\Phi_H\right)<0 .
 \label{general_superradiance_condition}
\end{equation}
For positive-frequency incident waves, \(\omega>0\), this reduces to
\begin{equation}
 0<\omega<q_s\Phi_H,
 \qquad
 \Phi_H=\frac{Q}{(1-\ell)r_+},
 \label{superradiancecondition}
\end{equation}
provided \(q_sQ>0\).  If \(q_sQ<0\), the electrostatic term has the wrong sign and no positive-frequency superradiant band exists in this gauge.  Equivalently, the horizon flux is proportional to \((\omega-q_s\Phi_H)|{\cal T}|^2\); amplification occurs precisely when the field deposits negative Killing energy into the black hole.  This is why charged superradiance is the wave counterpart of the electric Penrose process.

The LV parameter enters Eq.~\eqref{superradiancecondition} twice.  First, the factor \((1-\ell)^{-1}\) directly enhances the electric potential for fixed \(Q\) and \(r_+\).  Second, \(r_+\) itself depends on \(\ell\) through both \(\alpha\) and \(Q_{\rm eff}^2\).  The magnetic charge \(p\) changes the upper edge of the band only through this horizon shift, while changing the angular barrier through \(s=q_sp\).  Hence two black holes with nearly equal shadows can still have different superradiant thresholds if their electric and magnetic contributions to \(Q_{\rm eff}^2\) are distributed differently.

\begin{figure*}[tbhp]
\centering
\includegraphics[width=\textwidth]{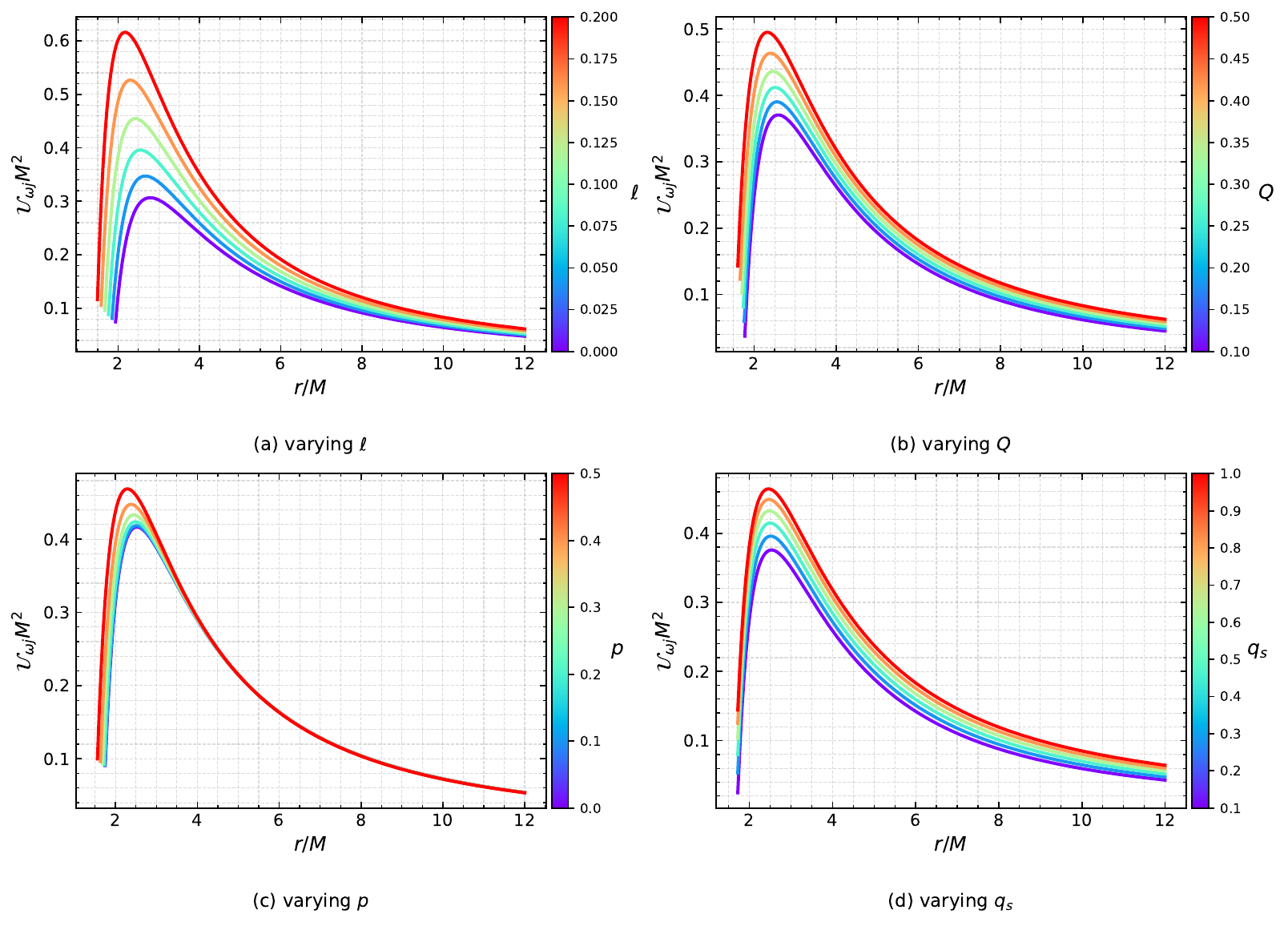}
\caption{Frequency-dependent charged-scalar barrier \(\mathcal U_{\omega j}(r)\) outside the horizon.  The benchmark values are \(M=1\), \(Q=0.30\), \(p=0.20\), \(q_s=0.55\), \(j=2\) and \(\omega M=0.45\), except for the parameter swept in each panel.  The color bars identify the varied parameter.}
\label{fig:chargedpotential}
\end{figure*}

Figure~\ref{fig:chargedpotential} displays the charged-scalar effective barrier in the same multi-panel style used for the other plots.  The barrier is plotted only outside the outer horizon.  Increasing \(\ell\), \(Q\), \(p\), or \(q_s\) changes not only the height of the barrier but also the radial interval available for scattering.  The panels therefore visualize two effects simultaneously: the deformation of the geometry and the electromagnetic tilting of the wave equation.

\begin{figure*}[tbhp]
\centering
\includegraphics[width=\textwidth]{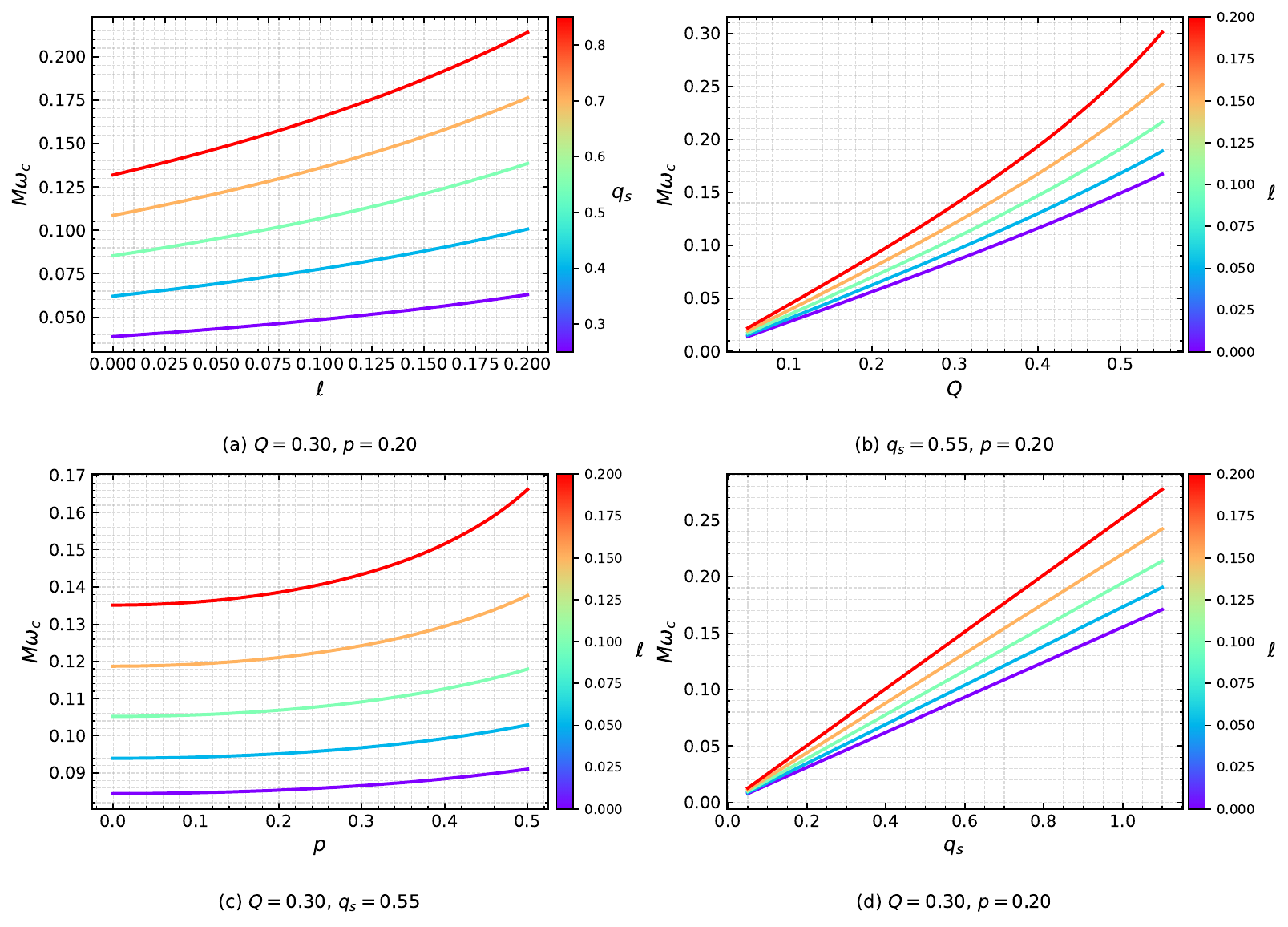}
\caption{Upper edge \(\omega_c=q_s\Phi_H\) of the charged-scalar superradiant window.  The four panels show how the threshold varies with the Lorentz-violating parameter, electric charge, magnetic charge and scalar charge.  Superradiant amplification is present for \(0<\omega<\omega_c\) when \(q_sQ>0\).}
\label{fig:superradiance}
\end{figure*}

Figure~\ref{fig:superradiance} shows the upper edge \(\omega_c=q_s\Phi_H\) of the superradiant window.  The threshold grows with \(\ell\) for the benchmark sequence because the horizon potential increases.  It also grows approximately linearly with the scalar charge \(q_s\), while the dependence on \(p\) is indirect and comes from the shift of \(r_+\).  These trends agree with the particle result that the critical charge \(q_c=E/\Phi_H\) decreases when \(\Phi_H\) increases.

With the analysis performed in this section, one observes that the effect of
superradiance by itself is not an instability.  In the asymptotically flat branch considered here, an amplified wave packet can escape to infinity after extracting electromagnetic energy from the black hole.  Exponential growth requires an additional trapping mechanism, such as a massive charged field, a mirror-like boundary condition, or anti-de Sitter asymptotics \cite{PressTeukolsky1972,FuruhashiNambu2004}.  Therefore, the dyonic Kalb-Ramond background supplies the amplification channel, but the existence of unstable bound states depends on the infrared boundary condition.  A complete instability analysis would require solving the corresponding quasi-bound-state problem with the appropriate normalization of time at infinity.

\section{Conclusions}\label{sec:conclusion}

In this work, we have analyzed charged-particle dynamics around a dyonic Kalb--Ramond black hole by examining the electric Penrose process, the magnetic Penrose process, high-energy particle collisions and the charged-scalar superradiant channel in the same Lorentz-violating geometry.  The solution is particularly useful for this purpose because the metric sees an effective dyonic charge, while the electric, magnetic and wave-extraction channels see the corresponding horizon potentials directly.

For a charged massive particle the conserved energy is $E=mF\dot t+q\Phi$,
while the future-directed kinetic quantity is $X=E-q\Phi$.  Negative conserved energy can occur outside the horizon only for $q\Phi<0$ and only sufficiently close to the horizon.  Thus the process is electromagnetic rather than ergospheric: the static geometry has no rotational ergoregion, but the canonical energy of an oppositely charged particle can be negative.  The lower energy bound is
\begin{equation}
    E\geq q\Phi(r)+\sqrt{F(r)}\sqrt{m^2+\frac{L^2}{r^2}},
\end{equation}
which approaches $q\Phi_H$ at the horizon. For a near-horizon decay in which fragment 1 falls into the black hole and fragment 2 escapes, the ideal electric Penrose efficiency satisfies $\eta\lesssim 1-\frac{q_1\Phi_H}{E_0}$, with
$\Phi_H=\frac{Q}{(1-\ell)r_+}$. For $q_1Q<0$, this bound exceeds unity.  Increasing the Lorentz-violating parameter raises the ideal extraction efficiency along the benchmark parameter path because it increases the horizon potential.  The magnetic charge modifies the electric efficiency indirectly by changing the horizon radius through $Q_{\rm eff}^2$.  When magnetically charged probes are included, however, there is an additional dual channel governed by
\begin{equation}
    \Psi_H=\frac{p}{(1-2\ell)r_+},
    \qquad
    E=mF\dot t+q\Phi+g\Psi .
\end{equation}
A negative magnetic fragment has $g\Psi_H<0$, and the ideal magnetic efficiency obeys $\eta_{\rm mag}\lesssim1-g_1\Psi_H/E_0$.  The full dyonic bound is $\eta_{\rm em}\lesssim1-(q_1\Phi_H+g_1\Psi_H)/E_0$, so electric and magnetic extraction can either reinforce or cancel each other depending on the signs of the absorbed fragment charges.  We also obtained the exact center-of-mass energy for two equatorial charged particles, namely
\begin{equation}
    E_{\rm cm}^2=m_1^2+m_2^2+2\left[\frac{X_1X_2-\sigma_1\sigma_2 Z_1Z_2}{F(r)}-\frac{L_1L_2}{r^2}\right],
\end{equation}
with $Z_i^2=X_i^2-F(m_i^2+L_i^2/r^2)$.  Same-direction collisions of two usual particles remain finite at a nonextremal horizon.  Head-on collisions can be formally divergent near any horizon if an outgoing particle exists there.  The charged BSW divergence for same-direction motion requires a critical particle satisfying $E=q\Phi_H$.  Such a particle cannot reach a nonextremal horizon, but it can reach an extremal horizon if the reachability condition
$E^2\geq \alpha\left(m^2+\frac{L^2}{r_e^2}\right)$ is satisfied.

The escape and censorship analyses add two important qualifications to this idealized efficiency picture.  First, a positive-energy product contributes to an observable Penrose event only if it lies above the outer effective barrier, namely $E_2\geq E_{\rm esc}$ with $E_{\rm esc}$ defined by Eq.~\eqref{Eesc_def}.  A large local collision energy is therefore not by itself sufficient for energy extraction at infinity.  Second, the absorbed fragment must leave the final parameters inside the horizon domain $\mathcal C_f\geq0$.  The possible dangerous interval $E_{\rm abs}\leq E<E_{\rm over}$ provides a useful test-particle diagnostic, but a definitive censorship statement requires the backreaction, charge normalization and second-order perturbation theory of the full Kalb-Ramond-electromagnetic system.  The monopole angular analysis also shows that generic dyonic products move on cones and that the radial barrier is governed by $J^2-(qp-gQ)^2$, not by the ordinary equatorial angular momentum alone.

This analysis shows us that charged-particle observables probe the dyonic Kalb-Ramond black hole differently from shadows, lensing and neutral scalar waves.  Metric observables mostly constrain $\alpha$ and $Q_{\rm eff}^2$, while electric energy extraction directly probes $Q/[(1-\ell)r_+]$ and magnetic extraction probes $p/[(1-2\ell)r_+]$.  This complementarity can help separate electric charge, magnetic charge and Lorentz violation, although significant degeneracies remain.  Future extensions should include numerical integration of dyonic charged-particle trajectories on monopole cones, collisional Penrose processes with full four-momentum conservation and escape cones, self-force estimates, second-order censorship tests, and the possible discharge of the black hole by electric and magnetic pair creation in the LV dyonic background.

\begin{acknowledgments}
The author Fernando Belchior would like to express gratitude to the Conselho Nacional de Desenvolvimento Cient\'{i}fico e Tecnol\'{o}gico CNPq for grant No. 151845/2025-5.
\end{acknowledgments}

\appendix

\section{Derivation of the radial equation}

Starting from the canonical energy $E=mF\dot t+q\Phi$, we define $X=E-q\Phi$.  Then $\dot t=X/(mF)$.  In the equatorial plane the angular momentum is $L=mr^2\dot\phi$.  The normalization condition gives
\begin{equation}
    -F\dot t^2+\frac{\dot r^2}{F}+r^2\dot\phi^2=-1.
\end{equation}
Substituting $\dot t$ and $\dot\phi$ gives
\begin{equation}
    -\frac{X^2}{m^2F}+\frac{\dot r^2}{F}+\frac{L^2}{m^2r^2}=-1.
\end{equation}
Multiplying by $F$ yields
\begin{equation}
    \dot r^2=\frac{X^2}{m^2}-F\left(1+\frac{L^2}{m^2r^2}\right),
\end{equation}
which is Eq.~\eqref{radial_dimful}.

\section{Center-of-mass energy formula}

For each particle, we have
\begin{equation}
    u_i^t=\frac{X_i}{m_iF},
    \qquad
    u_i^r=\frac{\sigma_i Z_i}{m_i},
    \qquad
    u_i^\phi=\frac{L_i}{m_ir^2}.
\end{equation}
Therefore, one obtains
\begin{align}
    -g_{\mu\nu}u_1^\mu u_2^\nu
    &=F u_1^t u_2^t-\frac{u_1^r u_2^r}{F}-r^2u_1^\phi u_2^\phi\nonumber\\
    &=\frac{X_1X_2-\sigma_1\sigma_2Z_1Z_2}{m_1m_2F}-\frac{L_1L_2}{m_1m_2r^2}.
\end{align}
Using $E_{\rm cm}^2=m_1^2+m_2^2+2m_1m_2\gamma$ gives Eq.~\eqref{Ecm_general}.

\section{Negative-energy boundary for radial motion}

For $L=0$ and $qQ<0$, the boundary of the negative-energy region is given by
\begin{equation}
    \frac{|qQ|}{(1-\ell)r}=m\sqrt{\alpha-\frac{2M}{r}+\frac{Q_{\rm eff}^2}{r^2}}.
\end{equation}
After squaring and multiplying by $r^2$, one obtains
\begin{equation}
    \frac{q^2Q^2}{(1-\ell)^2}=m^2\left(\alpha r^2-2Mr+Q_{\rm eff}^2\right).
\end{equation}
Hence, we have
\begin{equation}
    r_{\rm neg}^{\pm}=\frac{M\pm\sqrt{M^2-\alpha\left(Q_{\rm eff}^2-\frac{q^2Q^2}{m^2(1-\ell)^2}\right)}}{\alpha}.
\end{equation}
The outer boundary is $r_{\rm neg}^{+}$ if it is real and larger than $r_+$.  The negative-energy interval is then $r_+<r<r_{\rm neg}^{+}$.

\section{Extremal critical-particle reachability}

At extremality, we have
\begin{equation}
    F(r)=\alpha\left(1-\frac{r_e}{r}\right)^2,
    \qquad
    r_e=\frac{M}{\alpha}.
\end{equation}
For a critical particle, $E=qQ/[(1-\ell)r_e]$.  Near $r_e$,
\begin{equation}
    X=E-q\Phi(r)=\frac{qQ}{1-\ell}\left(\frac{1}{r_e}-\frac{1}{r}\right)
    =\frac{qQ}{(1-\ell)r_e^2}(r-r_e)+O[(r-r_e)^2].
\end{equation}
Meanwhile, one writes $FH$ as follows
\begin{equation}
    FH=\frac{\alpha}{r_e^2}(r-r_e)^2\left(m^2+\frac{L^2}{r_e^2}\right)+O[(r-r_e)^3].
\end{equation}
The condition $Z^2=X^2-FH\ge0$ gives
\begin{equation}
    \frac{q^2Q^2}{(1-\ell)^2r_e^2}\geq \alpha\left(m^2+\frac{L^2}{r_e^2}\right).
\end{equation}
Using the critical relation for $E$ yields Eq.~\eqref{reachability_ext_E}.

\section{Magnetic negative-energy boundary}

For completeness we give the derivation of Eq.~\eqref{mag_boundary_quad}.  In the pure magnetic radial sector one sets \(q=0\), \(L=0\), and
\begin{equation}
    E_{\rm min}^{\rm mag}(r)=g\Psi(r)+m\sqrt{F(r)} .
\end{equation}
The boundary of the negative-energy layer satisfies \(E_{\rm min}^{\rm mag}=0\).  For \(gp<0\), this condition is
\begin{equation}
    \frac{|gp|}{(1-2\ell)r}=m\sqrt{F(r)} .
\end{equation}
Using \(F(r)=\alpha-2M/r+Q_{\rm eff}^2/r^2\) and multiplying by \(r^2\) gives
\begin{equation}
    \frac{g^2p^2}{(1-2\ell)^2}=m^2\left(\alpha r^2-2Mr+Q_{\rm eff}^2\right),
\end{equation}
which is Eq.~\eqref{mag_boundary_quad}.  The outer root defines \(r_{{\rm neg},m}^{+}\) whenever it is real and larger than the event horizon.

\bibliographystyle{apsrev4-2}
\bibliography{references}

\end{document}